\documentclass{egpubl}
\usepackage{eg2024}
 
\JournalSubmission    % uncomment for submission to Computer Graphics Forum
\usepackage[T1]{fontenc}
\usepackage{dfadobe}  

\usepackage{cite}  % comment out for biblatex with backend=biber
\BibtexOrBiblatex
\electronicVersion
\PrintedOrElectronic
\ifpdf \usepackage[pdftex]{graphicx} \pdfcompresslevel=9
\else \usepackage[dvips]{graphicx} \fi

\usepackage{egweblnk} 
\usepackage{lineno}
\title[Flow Symmetrization for Parameterized Constrained Diffeomorphisms]%
{Flow Symmetrization for Parameterized Constrained Diffeomorphisms}

\usepackage{empheq}

\usepackage{breqn}
\usepackage{booktabs}
\usepackage{multirow}
\usepackage{amsmath, amssymb}
\begin{document}

\author[]{Aalok Gangopadhyay\thanks{indicates equal contribution}, Dwip Dalal$^\dagger$, Progyan Das$^\dagger$, Shanmuganathan Raman \\
{\parbox{\textwidth}{\centering 
        Indian Institute of Technology Gandhinagar, India\\
        \texttt{\{aalok, dwip.dalal, progyan.das, shanmuga\}@iitgn.ac.in}
       }
}}

% \author{Aalok\thanks{Your Institute Name, Address of the Institute. Email: aalok@iitgn.ac.in} \and
% Dwip\thanks{Your Institute Name, Address of the Institute. Email: dwip@iitgn.ac.in} \and
% Progyan\thanks{Your Institute Name, Address of the Institute. Email: progyan@iitgn.ac.in}}

% uncomment for using teaser
% \teaser{
%  \includegraphics[width=0.9\linewidth]{eg_new}
%  \centering
%   \caption{New EG Logo}
% \label{fig:teaser}
%}

\maketitle

% {\parbox{\textwidth}{\centering $^1$TU Darmstadt \& Fraunhofer IGD, Germany\\
%          $^2$Graz University of Technology, Institute of Computer Graphics and Knowledge Visualization, Austria
% %        $^2$ Another Department to illustrate the use in papers from authors
% %             with different affiliations
 
%        }
% }

%-------------------------------------------------------------------------
\begin{abstract}
   Diffeomorphisms play a crucial role while searching for shapes with fixed topological properties, allowing for smooth deformation of template shapes. Several approaches use diffeomorphism for shape search. However, these approaches employ only unconstrained diffeomorphisms. In this work, we develop Flow Symmetrization - a method to represent a parametric family of constrained diffeomorphisms that contain additional symmetry constraints such as periodicity, rotation equivariance, and transflection equivariance. Our representation is differentiable in nature, making it suitable for gradient-based optimization approaches for shape search. As these symmetry constraints naturally arise in tiling classes, our method is ideal for representing tile shapes belonging to any tiling class. To demonstrate the efficacy of our method, we design two frameworks for addressing the challenging problems of Escherization and Density Estimation. The first framework is dedicated to the Escherization problem, where we parameterize tile shapes belonging to different isohedral classes. Given a target shape, the template tile is deformed using gradient-based optimization to resemble the target shape. The second framework focuses on density estimation in identification spaces. By leveraging the inherent link between tiling theory and identification topology, we design constrained diffeomorphisms for the plane that result in unconstrained diffeomorphisms on the identification spaces. Specifically, we perform density estimation on identification spaces such as torus, sphere, Klein bottle, and projective plane. Through results and experiments, we demonstrate that our method obtains impressive results for Escherization on the Euclidean plane and density estimation on non-Euclidean identification spaces. Code and results: \href{https://dwipddalal.github.io/FlowSymmetry/}{https://dwipddalal.github.io/FlowSymmetry/} 
%-------------------------------------------------------------------------
%  ACM CCS 1998
%  (see https://www.acm.org/publications/computing-classification-system/1998)
% \begin{classification} % according to https://www.acm.org/publications/computing-classification-system/1998
% \CCScat{Computer Graphics}{I.3.3}{Picture/Image Generation}{Line and curve generation}
% \end{classification}
%-------------------------------------------------------------------------
%  ACM CCS 2012
   % (see https://www.acm.org/publications/class-2012)
%The tool at \url{http://dl.acm.org/ccs.cfm} can be used to generate
% CCS codes.
%Example:


\end{abstract}

\begin{figure}[!htb]
  \centering
  \includegraphics[width=\linewidth]{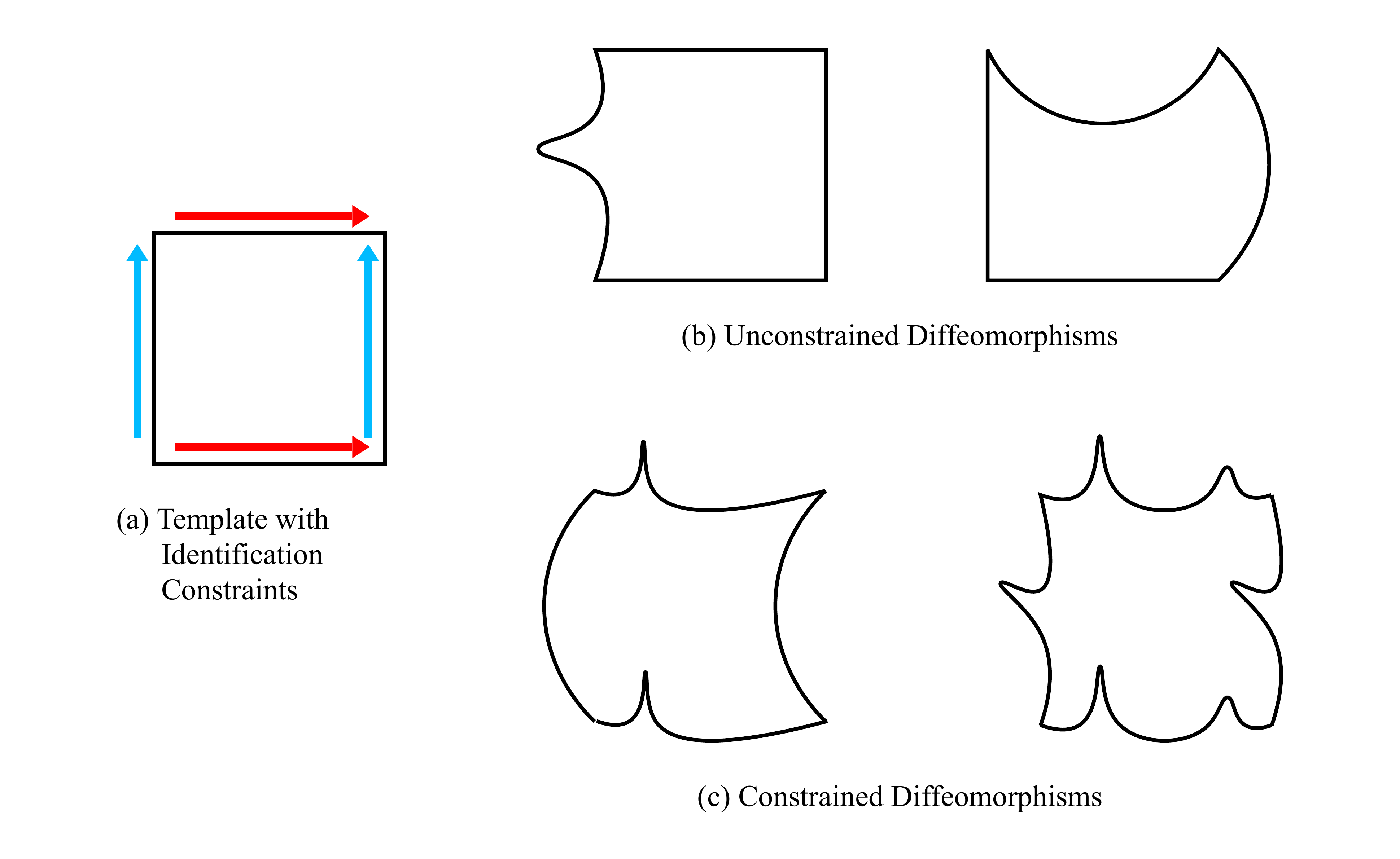}
  \caption{(a) Consider a template square with its opposite sides identified. The first pair of opposite sides are marked in blue, and the second one in red. The requirement is that while deforming this template square, the opposite sides must deform identically. (b) Using an unconstrained diffeomorphism to deform the square violates the required constraint. (c) Our flow symmetrization approach helps design diffeomorphisms where the template can be deformed while satisfying such identification constraints.}
  \label{fig:ConstrainedDiffeomorphism}
\end{figure}

%-------------------------------------------------------------------------

\section{Introduction}
\label{sec:Introduction}
%--------------------------------------------------%
% What is diffeomorphism? (bijective, bismooth)
% Why is it useful? (preserves toplogy)
% Different applications and problems where Diffs play an important role in Graphics
% Cite the successful works, both older and newer ones. 
% Refer to the google doc on LitRev
%--------------------------------------------------%

Shape search is a fundamental problem centered on finding a shape that satisfies specific constraints while optimizing a predefined objective. When the shape's topological properties are predetermined, diffeomorphisms become crucial. Diffeomorphisms are smooth transformations with smooth inverses. They thus facilitate the smooth deformation of a template shape, ensuring that its topological properties remain preserved. Such utility of diffeomorphisms has been effectively highlighted in \cite{sun2022topologypreserving, gupta2020neural, cheng2022diffeomorphic}.

%--------------------------------------------------%
% Different ways of representing diffeomorphisms
% Recent methods to model diffeos - INNs, Flows, NeuralODE
% advantages and disadvantages of different methods
% Introduction to flow based methods/ Neural ODEs 
% advantages of neuralODE based flows - differentiable (gradient-descent based optimization), query based (memory efficient)
%--------------------------------------------------%

Various methods are utilized to represent diffeomorphisms, including Invertible Neural Networks (INN) \cite{dinh2016density} and NeuralODEs \cite{chen2018neural}. Within the INN framework, a neural network is inherently designed to be smooth and invertible, ensuring its representation as a diffeomorphism. On the other hand, NeuralODEs allow for the representation of diffeomorphisms via stationary or time-varying vector fields. By integrating a given vector field, a corresponding diffeomorphism can be obtained. Both these methods present distinct advantages. Primarily, their differentiable nature renders them compatible with gradient-based optimization methods. Additionally, given that these models are query-based, the memory requirement scales with the complexity of the diffeomorphism, marking them as a memory-efficient solution.

%--------------------------------------------------%
% Limitation - the above methods are unconstrained diffeomorphism 
% Motivate through illustration the need for constrained diffeomorphisms
% what are constrainted diffeomorphisms
% When do we need them and why they can be useful
% Motivate with tiling and Identification topology (fundamental polygon) the need for the three symmetries in flow of the diffemorphisms - translation invariance (periodicity), rotational equivariance, transflection equivariance (glide-reflection)
%--------------------------------------------------%

While a rich body of work in the field successfully models diffeomorphisms, most of this literature focuses on unconstrained diffeomorphisms. These unconstrained representations, while powerful, do not always suffice when the problem contains certain constraints that need to be satisfied. Take, for example, the square depicted in Fig. \ref{fig:ConstrainedDiffeomorphism}(a) with its identification constraints. In this example, the opposite sides of the squares are identified so that the opposite sides must deform identically. Applying an unconstrained diffeomorphism on this square might violate these constraints, as seen in Fig. \ref{fig:ConstrainedDiffeomorphism}(b). To maintain these identification constraints while deforming, like the ones shown in Fig. \ref{fig:ConstrainedDiffeomorphism}(c), the diffeomorphism must possess symmetry attributes that ensure the constraints are not violated. In such contexts, the capability to employ constrained or symmetric diffeomorphisms becomes paramount, ensuring the retention of the shape's intrinsic symmetries.

\begin{figure}[!htb]
  \centering
  \includegraphics[width=0.85\linewidth]{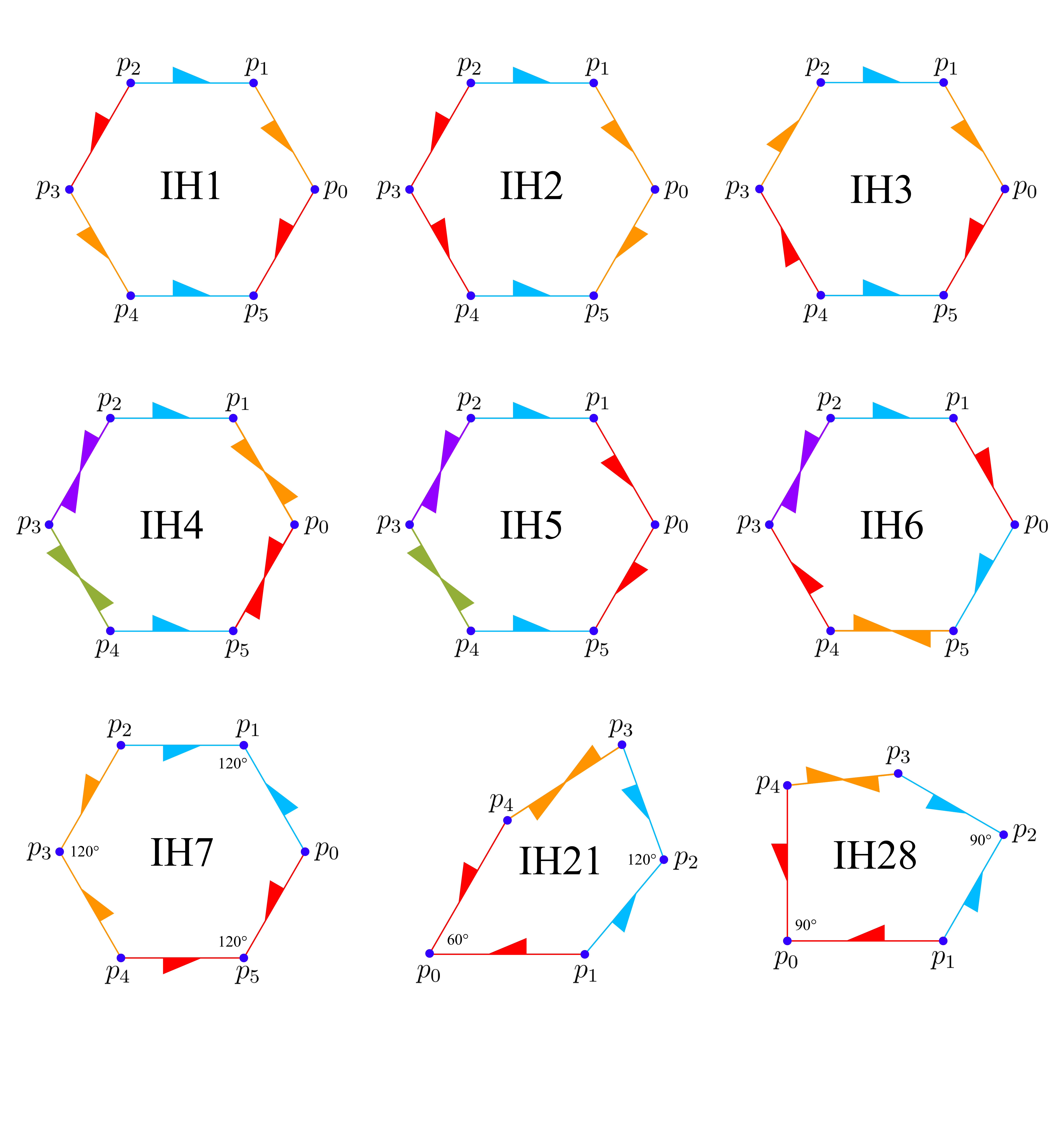}
  \caption{The identification constraints involved in 9 isohedral classes are shown. The identified edges are color-coded along with their orientation marked with an arrowhead. In some cases, there are additional angular constraints.}
  \label{fig:IsohedralClasses}
\end{figure}

\begin{figure}[!t]
  \centering
  \includegraphics[width=\linewidth]{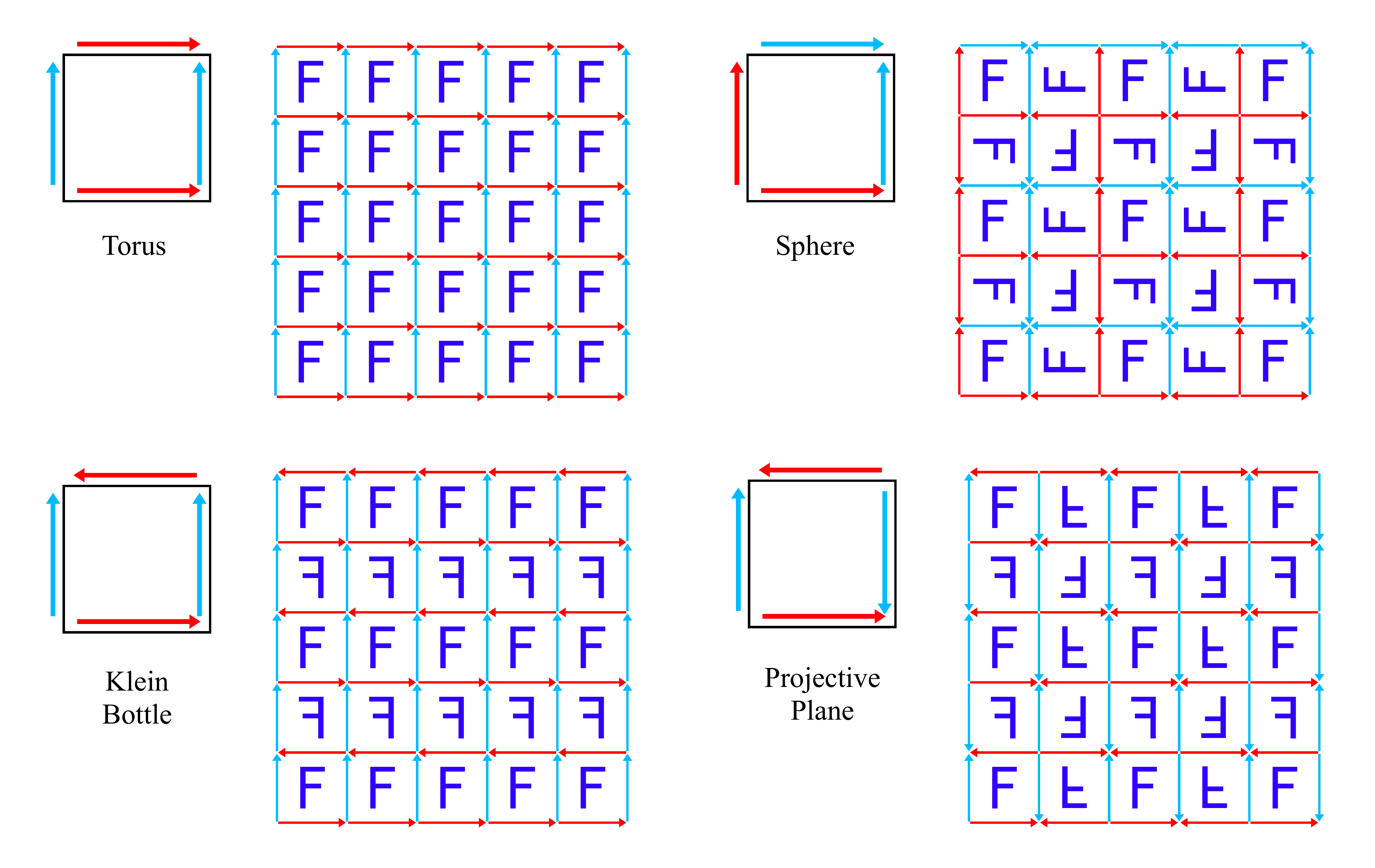}
  \caption{Different ways of identification in a template square result in different topological spaces such as torus, sphere, Klein bottle, and projective plane. Moreover, taking countable copies of the template square and joining them edge-to-edge based on the identification rule results in specific tiling patterns. The resulting tiling for each of the topological spaces is shown.}
  \label{fig:Identification}
\end{figure}

Identification constraints in polygons under certain conditions result in tiling patterns. Some examples can be seen in Fig. \ref{fig:IsohedralClasses} and Fig. \ref{fig:Identification}. The identification constraints in 9 isohedral tiling classes are displayed in Fig. \ref{fig:IsohedralClasses}. The template shapes are hexagons and pentagons with identification constraints on their edges. Taking countable copies of these templates and joining them according to the identification rule results in tiling patterns as shown in Fig. \ref{fig:IsohedralSymmetries}. These tiling patterns have one or more translation, rotation, and transflection symmetries in them. Similarly, in Fig. \ref{fig:Identification}, we see that the template square identified in different ways results in different topological spaces, each with its own tiling pattern containing symmetries.

Recognizing the importance of constrained diffeomorphisms, we present a novel methodology to represent diffeomorphisms with inherent symmetry constraints. These constraints stem from the identification pattern of the template polygons.
The representation of the diffeomorphism utilizes the NeuralODE framework. In this context, integrating over a static vector field results in a diffeomorphism. To ensure that this diffeomorphism abides by the required constraints, it's necessary to embed the desired symmetries directly into this vector field.
Three primary symmetries – periodic, rotational, and transflectional – emerge prominently in tiling patterns. Our method ensures these symmetries by enforcing periodicity, rotation equivariance, and transflection equivariance in the vector field.
Our main contribution is the introduction of a 'Flow Symmetrization' process. This process transforms an initially unconstrained diffeomorphism into one that respects the required symmetry constraints by symmetrizing the vector field accordingly.
% Recognizing the need for constrained diffeomorphism, we introduce a novel methodology to design diffeomorphisms that possess symmetry constraints arising from identification in the template polygons. We use the idea of NeuralODE, where integrating over a static vector field yields a diffeomorphism. Making the diffeomorphism constrained is equivalent to introducing corresponding symmetries in the static vector field. We show that the periodic, rotation and transflection symmetries arising in the tiling pattern can be ensured by introducing periodicity, rotation equivariance and transflection equivariance in the vector field. In this work, we develop a process of symmetrization where we convert an unconstrained diffeomorphism into a constrained diffeomorphism by enforcing the required symmetries in the vector field. We call this process as Flow Symmetrization, as it performs the required symmetrization process on the vector flow field.
We further delve into the practical ramifications of our Flow Symmetrization method by exploring two distinct applications: Escherization on the Euclidean plane and Density estimation on the identification spaces.

%--------------------------------------------------%
% Escherization
%--------------------------------------------------%
% 

In the Escherization problem, given a target shape, the objective is to find a shape that resembles the target shape and tiles the entire Euclidean plane. In our approach, we take the template tiles belonging to the nine isohedral classes (shown in Fig. \ref{fig:IsohedralClasses}) and deform them using constrained diffeomorphisms constructed using our Flow Symmetrization approach. For each isohedral class, we identify their symmetries and design a family of vector fields that contain those symmetries, yielding us constrained diffeomorphisms. Our method guarantees that the shapes obtained after deforming the template indeed tile the entire plane. We develop loss functions and use gradient-based optimization techniques to obtain the tile shape closest to the target shape. This is possible owing to the fact that our approach is fully differentiable. Moreover, our method works for non-polygonal target shapes and can generate non-polygonal tiles, thus overcoming a limitation of the previous approaches.

%--------------------------------------------------%
% Density Estimation
%--------------------------------------------------%

In the density estimation problem, given samples from an unknown target distribution, the objective is to find a distribution that is close to the target. In our approach, we perform density estimation on identification spaces. We achieve this by designing diffeomorphisms on these spaces. Identification spaces considered here arise from identification constraints on the square, which also gives rise to a tiling pattern. We designed a diffeomorphism on the Euclidean plane that satisfied the symmetries of the identification. The constrained diffeomorphism combined with a canonical projection yields a diffeomorphism on the required identification space, which enables us to perform density estimation.

The following are the major contributions of this work:
\begin{enumerate}
    \item We develop a differentiable parametric representation of constrained diffeomorphisms for the deformation of templates under identification constraints. 
    \item We introduce a Flow Symmetrization method for enforcing periodicity, rotation equivariance, and transflection equivariance in the vector flow field.
    \item We develop a method to parametrize tile shapes belonging to isohedral tiling classes. Using this we design a gradient-based optimization framework for the Escherization problem to find the tile shape resembling a given target shape. We also develop a novel loss function to facilitate the optimization process.
    \item By utilizing the diffeomorphisms on Euclidean spaces and enforcing identification constraints in them, we are able to design diffeomorphisms on identification spaces such as torus, sphere, Klein bottle, and projective plane. We further use this diffeomorphic flow to perform density estimation on these identification spaces.
\end{enumerate}

To the best of our knowledge, we are the first to develop a differentiable parametric representation of diffeomorphisms with identification constraints to solve Escherization on the plane and density estimation on identification spaces using a gradient-based optimization framework. 
% NOTE: "We are the first" instead of "first ones".

\begin{figure*}[!htb]
  \centering
  \includegraphics[width=\linewidth]{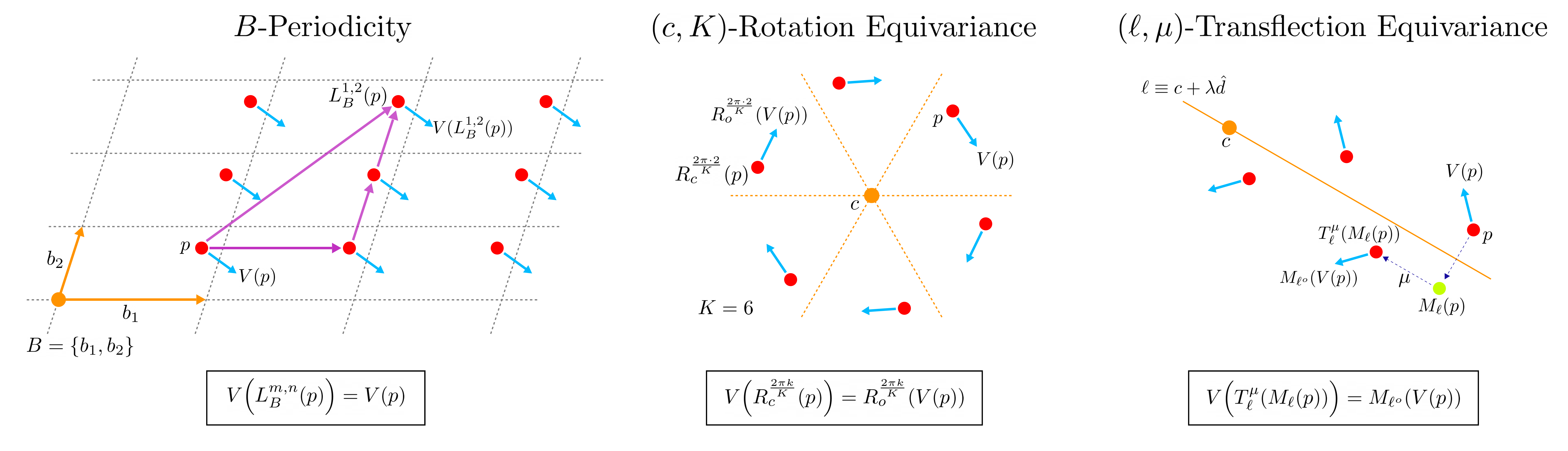}
  \caption{Three types of symmetry constraints arise in the underlying vector field of the diffeomorphism under identification constraints: periodicity, rotation equivariance, and transflection equivariance. These symmetries are illustrated, and their corresponding equations of invariance or equivariance in the vector field are highlighted.}
  \label{fig:SymmetryConstraints}
\end{figure*}

\section{Background}
\label{sec:Background}

% \begin{table}[!tbh]
% \centering
% \begin{tabular}{cc}
% \toprule
% \textbf{Notation} & \textbf{Name/Description} \\
% \hline
% $V: \mathbb{R}^2 \rightarrow \mathbb{R}^2$ & Vector Field \\
% $D: \mathbb{R}^2 \rightarrow \mathbb{R}^2$ & Diffeomorphism \\
% $\theta \in \Theta$ & Parameter space \\
% $V_\theta, D_\theta$ & Parameterized family \\
% $t, t_0, t_1$ & Time \\
% $s(t)=(s_1(t), s_2(t))$ & Point on plane \\
% $p, q$ & points on plane\\
% $B=\{b_1, b_2\}$ & Basis for Lattice \\
% $n, i, j, k$ & dummy integers \\
% $\Omega_1, \Omega$ & Constant integers \\
% \bottomrule
% \end{tabular}
% \end{table}

We first describe the representation of unconstrained diffeomorphisms using integration over static vector fields. We then describe the symmetry constraints arising in diffeomorphisms with identification constraints.

\subsection{Diffeomorphism}
\label{sec:diffNeuralODE}

% Flow is defined as a mapping, 

% $$\varphi: X \times \mathbb{R} \rightarrow X, $$ 
% from a position, $x \in X$ and time interval $t \in\mathbb{R}$, to a different position $x' \in X$ -- in effect, we consider that a point at $x$ has ``flowed'' in time $t$ from $x$ to $x'$. Hence, over two time intervals $s$ and $t$, and a position $x$, we can always say,

% \begin{empheq}{align}
% & \varphi(x, 0)=x \text{  (identity property)} \nonumber\\
% & \varphi(\varphi(x, t), s)=\varphi(x, s+t) .
% \label{eq:Flow}
% \end{empheq}

% In the case of flows over manifolds defined by a differential equation, we may consider the $x$ to be dependent on time and the initial condition, $x = x_0$. We can, therefore, consider the \textit{flow} as a series of invertible, parameterized maps that can be learnt.

%--------------------------------------------------%
% Define Vector fields
% IVP 
% Trajectories and integral curves
% Picard Lindelof thm
% Diffeomorphism - bijective, bismooth
% Diffeo as integration of a flow
%--------------------------------------------------%

Consider a vector field $V: \mathbb{R}^2 \to \mathbb{R}^2$.
This gives rise to the flow of the form
\begin{empheq}{align}
% \begin{empheq}[box=\fbox]{align}
\frac{ds(t)}{dt}    & = V(s(t)) \\ 
 & s(t_0) = q \nonumber
\label{eq:IVP} 
\end{empheq}
where, $t \in \mathbb{R}$ denotes time, $s(t) \in \mathbb{R}^2$ denotes a point on the euclidean plane,  $t_0$ denotes initial time and $q \in \mathbb{R}^2$ denotes the initial point.
Further, assume that the vector field $V$ is smooth. Under this assumption, based on the Picard–Lindel\"{o}f theorem, the initial value problem has a unique solution. 
This gives rise to a diffeomorphism $D: \mathbb{R}^2 \to \mathbb{R}^2$ given by 
\begin{empheq}{align}
% \begin{empheq}[box=\fbox]{align}
D(q) = s(t_1) = s(t_0) + \int_{t_0}^{t_1} V(s(t)) \,dt
\label{eq:DiffeoIVP} 
\end{empheq}

We call $D$ to be the diffeomorphism induced by the vector field $V$ and call $V$ to be the underlying vector field of $D$. Moreover, having a family of vector fields $V_{\theta}$ parameterized by $\theta \in \Theta$ gives rise to a parametrized family of diffeomorphisms $D_{\theta}$. We want to construct a parametrized representation of diffeomorphisms with symmetry constraints in their underlying vector fields.

% Consider a diffeomorphism $\mathcal{D}: \mathbb{R}^2 \to \mathbb{R}^2$ defined on the Euclidean plane. This implies that $\mathcal{D}$ is a bijective and smooth mapping with a smooth inverse, denoted as $\mathcal{D}^{-1}$.
% Consider a flow field $V_{\theta}:\mathbb{R}^2 \times \mathbb{R} \to \mathbb{R}^2$ parameterized by $\theta \in \Theta$.
% Let us consider an initial value problem of the form 
% \begin{empheq}{align}
% \begin{empheq}[box=\fbox]{align}
% \frac{dz(t)}{dt}    & = V_{\theta}(z(t)) \\ 
%  & z(t_0) = q \nonumber
% \label{eq:IVP} 
% \end{empheq}
% where, $t \in \mathbb{R}$ denotes time, $z(t) \in \mathbb{R}^2$ denotes a point on the euclidean plane,  $t_0$ denotes initial time and $q \in \mathbb{R}^2$ denotes the initial point. 
% Further assume that the flow field $V_{\theta}$ is smooth in both $z$ and $t$. Under this assumption, based on the Picard–Lindelöf theorem, the initial value problem has a unique solution. 
% This gives rise to a family of diffeomorphisms $\mathcal{D}_{\theta}: \mathbb{R}^2 \to \mathbb{R}^2$ parameterized by $\theta \in \Theta$ given by 
% \begin{empheq}{align}
% % \begin{empheq}[box=\fbox]{align}
% \mathcal{D}_{\theta}(q) = z(t_1) = z(t_0) + \int_{t_0}^{t_1} V_{\theta}(z(t),t) \,dt
% \label{eq:DiffeoIVP} 
% \end{empheq}

\subsection{Constrained Diffeomorphism}
%--------------------------------------------------%
% Constrained Diffeomorphism (mathematically formulate the constraints in the flow of a diffeo)
% >> Translation invariance 
% >> Rotational equivariance
% >> Transflection equivariance
%--------------------------------------------------%

Identification constraints on the edges of the fundamental polygon give rise to three types of symmetry constraints on the underlying vector flow field of the diffeomorphism. We define each of these symmetry constraints in this section. These symmetries are illustrated in Fig. \ref{fig:SymmetryConstraints}

\subsubsection{$B$-periodic flow}
Consider  $B=[b_1,b_2]$ a basis matrix in $\mathbb{R}^2$. This implies that $b_1$ and $b_2$ are linearly independent vectors, and together, they span the entire Euclidean plane, $\mathbb{R}^2$.
Let  $p \in \mathbb{R}^2$ be a point on the plane and $m,n \in \mathbb{Z}$ be integers.
Let us define $L_B^{m,n}(p) = p + m b_1 + n b_2$, where $L_B^{m,n}(p)$ denotes the translation of a point $p \in \mathbb{R}^2$ by $(m,n)$ units along the lattice generated by $B$.
%%% Need to restructure this by introducing the notion of a cell and lattice terminologies
A vector flow field $V$ is said to be $B$-periodic if, $\forall p \in \mathbb{R}^2$ and $\forall m,n \in \mathbb{Z} $, we have
\begin{empheq}{align}
% \begin{empheq}[box=\fbox]{align}
V\Big(L_B^{m,n}(p)\Big) = V(p) 
\label{eq:PeriodicFlow} 
\end{empheq}

\subsubsection{$(c,K)$-rotation equivariant flow}
Let $o=(0,0)$ denote the origin and let $c \in \mathbb{R}^2$ be a point having $K$-fold rotational symmetry. Let $R_{o}^{\theta}$ and $R_{c}^{\theta}$ denote counter-clockwise rotation by an angle $\theta$ about $o$ and $c$, respectively. 
A vector field $V$ is said to be $(c,K)$-Rotation equivariant if $\forall k \in \{0,1,\cdots,K-1\}$, we have,
\begin{empheq}{align}
% \begin{empheq}[box=\fbox]{align}
V\Big(R_{c}^{\frac{2\pi k}{K}}(p)\Big) = R_{o}^{\frac{2\pi k}{K}}(V(p))
\label{eq:RotEqFlow} 
\end{empheq}

\subsubsection{$(\ell,\mu)$-transflection equivariant flow}
Consider the line $\ell \equiv c + \lambda \hat{d}$. The mirror reflection of a point $p$ about $\ell$ is given by $M_{\ell}(p) = 2c - p + 2\langle p-c,\hat{d} \rangle\hat{d}$. The translation of a point $p$ by a distance $\mu$ parallel to the line $\ell$ is given by $T_{\ell}^{\mu}(p) = p + \mu \hat{d}$. The composition of reflection followed by translation results in a transflection $G_{\ell}^{\mu}(p) = T_{\ell}^{\mu}(M_{\ell}(p))$. Let $\ell^{o} \equiv \lambda \hat{d}$ be the line passing through the origin and parallel to $\ell$. A vector field $V$ is said to be $(\ell,\mu)$-transflection equivariant if
\begin{empheq}{align}
% \begin{empheq}[box=\fbox]{align}
V\Big(T_{\ell}^{\mu}(M_{\ell}(p))\Big) = M_{\ell^{o}}(V(p)) 
\label{eq:TrflcEqFlow} 
\end{empheq}

\section{Flow Symmetrization}
\label{sec:Symmetrization}
% How the three symmetries are enforced in out approach
% In the tiling wallpaper groups, periodicity is present in all. Additionally it might include rotation symmetry and or transflection symmetry.

% Explain that we use the fourier series defined on the lattice with unit square. And for an arbitrary lattice, we change the basis, with a linear transformation.

The idea of flow symmetrization is as follows: Given an arbitrary vector field and a set of symmetry constraints, we apply the symmetrization process, resulting in a vector field satisfying those required symmetries. We would require $B$-periodic vector fields with or without rotation equivariances and transflection equivariances for the applications mentioned in this paper.
In this section, we first mention how to construct $B$-periodic vector fields for arbitrary $B$. We then explain how an arbitrary vector field is made $(c, K)$-Rotation equivariant or $(\ell,\mu)$-transflection equivariant. 

First, we construct a periodic vector field for the canonical basis $B_S = [(1,0),(0,1)]$ representing a unit square. Let $\Theta = \mathbb{R}^{2 \times 4 \times \Omega_1 \times \Omega_2}$ be the parameter space. Let $\theta_{i,j}^{k,l} \in \Theta$ be a particular choice of parameter with $k \in \{1,2\}$ indicating the coordinate index and $l \in \{1,2,3,4\}$ indicating the term of the 2-dimensional Fourier series. Here $i \in \{0,1,\cdots,\Omega_1\}$ and $j \in \{0,1,\cdots,\Omega_2\}$ denote the frequency multiplier of the first and second coordinates, respectively. For a point $p=(x,y)$,
the periodic flow with the unit square as a translation unit is given by 
\begin{empheq}{align}
F(p) = F((x,y)) = (f(x,y,1),f(x,y,2))
\label{eq:F_function}
\end{empheq}
where $f(x,y,k)$ is given by
\begin{empheq}{align}
% \begin{empheq}[box=\fbox]{align}
f(x,y,k) = & \sum\limits_{i=0}^{\Omega_1}\sum\limits_{j=0}^{\Omega_2} \theta_{i,j}^{k,1} cos(2 \pi i x ) cos(2 \pi j y ) \nonumber \\
& \sum\limits_{i=0}^{\Omega_1}\sum\limits_{j=0}^{\Omega_2} \theta_{i,j}^{k,2} cos(2 \pi i x ) sin(2 \pi j y ) \nonumber \\
& \sum\limits_{i=0}^{\Omega_1}\sum\limits_{j=0}^{\Omega_2} \theta_{i,j}^{k,3} sin(2 \pi i x ) cos(2 \pi j y ) \nonumber \\
& \sum\limits_{i=0}^{\Omega_1}\sum\limits_{j=0}^{\Omega_2} \theta_{i,j}^{k,4} sin(2 \pi i x ) sin(2 \pi j y ) 
\label{eq:Fourier} 
\end{empheq}

For an arbitrary lattice basis B, we obtain a $B$-periodic vector field by using the periodic vector field on the unit square with a change of basis operation given by
\begin{empheq}{align}
V(p) = B(F(B^{-1}(p)))
\end{empheq}

For an arbitrary vector field $V_0$, a $(c, K)$-rotation equivariant vector field is obtained through the following symmetrization equation.
\begin{empheq}{align}
V(p) = \sum_{k=1}^{K} R_{o}^{-\frac{2\pi k}{K}} V_0\Big(R_{c}^{\frac{2\pi k}{K}}(p)\Bigg)
\end{empheq}

Given an arbitrary vector field $V_0$, we obtain a $(\ell,\mu)$-transflection equivariant vector field through the following symmetrization equation.

\begin{empheq}{align}
V(p) = V_0(p) +
M_{\ell^{o}}\Big(V_0\Big(T_{\ell}^{\mu}(M_{\ell}(p))\Big)\Big)
\end{empheq}
This can be compactly written as
\begin{empheq}{align}
V(p) = \sum_{k=0}^{1}M_{\ell^{o}}^{k}\Big(V_0\Big({T_{\ell}^{\mu}}^{k}(M_{\ell}^{k}(p))\Big)\Big) 
\end{empheq}
where, for $k=0$, the functions $M_{\ell^{o}}^{k}$, ${T_{\ell}^{\mu}}^{k}$ and $M_{\ell}^{k}$ are all identity.

For convenience of notation, we introduce the symmetrization operators for the three types of symmetries.
The basis symmetrization operator, the rotation symmetrization operator, and the transflection symmetrization operator are given in Eq. \ref{eq:basisSymmOp}, Eq. \ref{eq:rotationSymmOp} and Eq. \ref{eq:transflectionSymmOp}, respectively.

% The basis symmetrization operator is given by
\begin{empheq}{align}
\mathcal{P}_B(V) = B \circ V \circ B^{-1}
\label{eq:basisSymmOp}
\end{empheq}

% The rotation symmetrization operator is given by
\begin{empheq}{align}
\mathcal{R}_c^K(V) = \sum_{k=1}^{K} R_{o}^{-\frac{2\pi k}{K}} \circ V \circ R_{c}^{\frac{2\pi k}{K}}
\label{eq:rotationSymmOp}
\end{empheq}

% The transflection symmetrization operator is given by
\begin{empheq}{align}
\mathcal{G}_{\ell}^{\mu}(V) = \sum_{k=0}^{1}M_{\ell^{o}}^{k} \circ V \circ {T_{\ell}^{\mu}}^{k} \circ M_{\ell}^{k} 
\label{eq:transflectionSymmOp}
\end{empheq}

These symmetrization operators would be used for constructing symmetric vector fields for the problems of Escherization (Sec. \ref{sec:Escherization}) and density estimation (Sec. \ref{sec:FlowOnManifolds}).

\section{Escherization}
\label{sec:Escherization}

%--------------------------------------------------%
% Escherization
% Tiling basics, Isohedral classes (for details, refer to Esch1, Esch2, Graunbaum, and Shepard)
%--------------------------------------------------%

 Tiling, an integral discipline of geometric study, concerns the assembly of shapes in a manner that completely fills a plane without any gaps or overlaps. A well-studied problem within this field is the problem of Escherization, formally defined as:
 
 \hspace{2pt}

\fbox{
\begin{minipage}{0.42\textwidth}
Given an input shape $\mathcal{S} \subseteq \mathbb{R}^2$, where $\mathcal{S}$ is topologically a closed disc, find $\mathcal{S}^{\prime}$, also of closed disc topology, such that countable copies of $\mathcal{S}^{\prime}$ tile $\mathbb{R}^2$, and, $d\left(\mathcal{S}, \mathcal{S}^{\prime}\right)$ is minimized, for a given metric $d$
\end{minipage}
}

 \hspace{2pt}

The problem of Escherization has been traditionally addressed through the use of isohedral tiling, a class of tiling characterized by a single tile shape and translation symmetry, where a configuration of one or more adjacent tiles repeats periodically across the entire tiling \cite{kaplan2000escherization, koizumi2011maximum, nagata2021escherization}. This tiling class offers the flexibility to create a wide range of tiling patterns. For each isohedral type, any conceivable tile shape can be derived by deforming a polygonal template within certain constraints, which typically has no more than six vertices. The isohedral tiling class is subdivided into 93 distinct groups, labeled IH1 through IH93, each differentiated based on the adjacency relations between tiles \cite{grunbaum1987tilings}.

Isohedral tilings present a consistent structural framework for analyzing translational symmetry in geometric configurations. Each class, denoted by its unique identifier, provides specific insights into geometric and topological properties. When the adjacency relationships between tiles are ignored, every conceivable tile shape can be generated using the templates of the nine primary isohedral types \cite{schattschneider2004escher}, as depicted in Fig. \ref{fig:IsohedralClasses}. 

\begin{table}[t]
\resizebox{\linewidth}{!}{
\begin{tabular}{l|lllllllll}
\toprule
Type & IH1 & IH2 & IH3 & IH4 & IH5 & IH6 & IH7 & IH21 & IH28 \\
\multirow{2}{*}{Symmetry} & \multicolumn{9}{c}{} \\
\multirow{2}{*}{Group} & p1  & pg  & pg  & p2  & pgg & pgg & p3  & p6   & p4  \\
\addlinespace[0.5em]
\bottomrule
\end{tabular}
}
\caption{9 types of IH classes and their corresponding symmetry groups.}
\label{tab: 1}
\end{table}

Each IH class has an associated symmetry group (Tab. \ref{tab: 1}) that defines the nature of the tiling pattern in that group and indicates the adjacency relation of tiles in the tiling pattern. For example, the IH1 class corresponds to the simplest \texttt{p1} symmetry group, which involves only translations. On the other hand, the IH5 and IH6 classes are associated with the \texttt{pgg} symmetry group, which includes glide-reflections and 2-fold rotations but no reflections (details of each of these symmetry groups can be found in the supplementary material). These symmetry groups serve as the mathematical backbone for the different types of isohedral tilings, providing a structured way to represent them. We leverage these nine isohedral types and their symmetries for solving the Escherization problem. We tailor our method by enforcing these symmetry constraints using our Flow Symmetrization approach, ensuring that all the template tile deformation due to the constrained diffeomorphisms would preserve the IH class symmetry, thereby generating all potential tile shapes. This allows us to address the Escherization problem for generating arbitrary tile shapes that need not be restricted to polygons.

\subsection{Methodology}

\begin{figure}[!t]
  \centering
  \includegraphics[width=\linewidth]{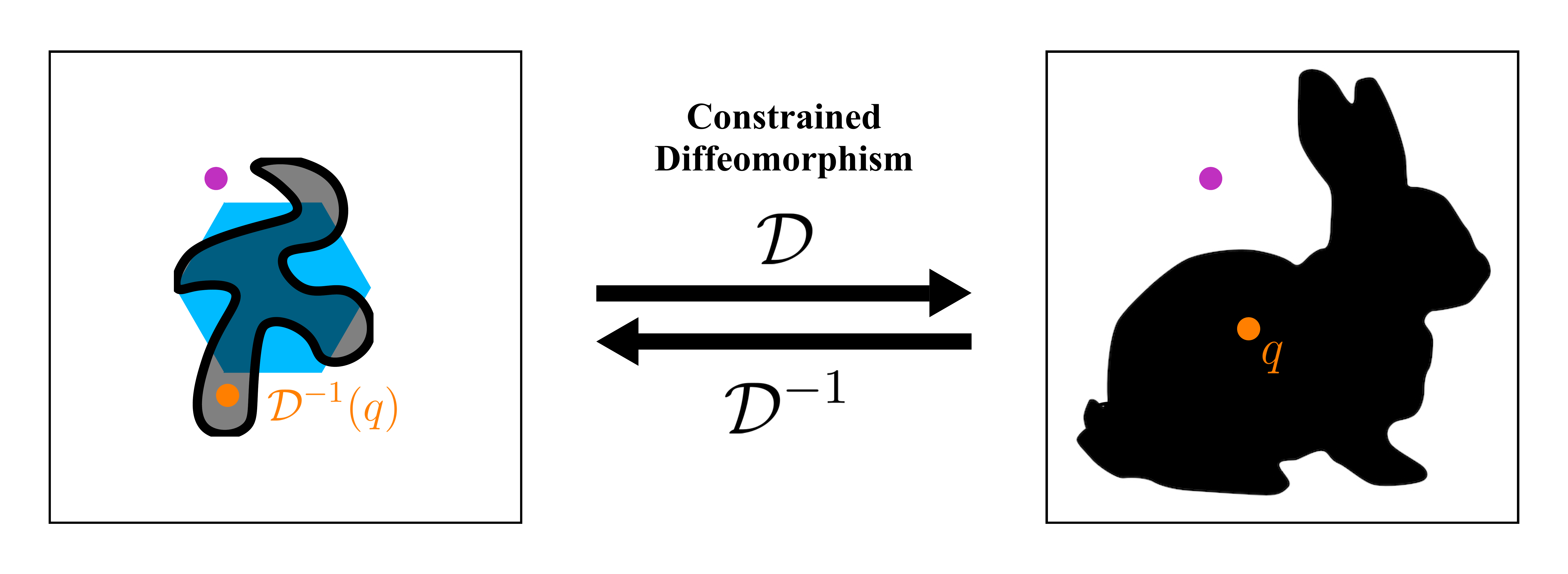}
  \caption{In our approach, a template tile $T$ belonging to an isohedral class is deformed using a constrained diffeomorphism $D$ to resemble a target shape $S$, such as a rabbit. To find the optimal diffeomorphism, however, we use it in inverse mode. We sample points from the domain of the target shape. These points are mapped to the template domain through $D^{-1}$. After the inverse mapping, points initially inside $S$ must lie inside $T$, and points initially outside $S$ must lie outside $T$. Sampled points that do not obey this property are penalized using the occupancy loss. Thus, using a gradient-based optimization technique, our differentiable framework helps find the ideal diffeomorphism by minimizing the occupancy loss. This diffeomorphism when applied on template $T$, yields the tile shape resembling $S$, thus solving the Escherization problem.}
  \label{fig:Methodology}
\end{figure}

To address the task of Escherization for general tile shapes that are not restricted to polygons, we integrate the concept of constrained diffeomorphisms, as introduced in Section 3, for tile shape deformations. Specifically, we employ Neural Ordinary Differential Equations (Neural ODEs) \cite{chen2018neural} to generate diffeomorphisms as explained in Sec. \ref{sec:diffNeuralODE}. The vector flow fields of these diffeomorphisms are then subjected to symmetry constraints based on the isohedral (IH) class to which the tile shape belongs. These constrained diffeomorphisms are then applied to the template tile of the IH classes to generate arbitrary tile shapes. The template shape is hexagonal for IH classes IH1 to IH7 and pentagonal for IH21 and IH28. These symmetry constraints guarantee that the deformed shape indeed tiles the entire plane, and we try to find among these tiles the one that is closest to the target shape, thus fulfilling the Escherization criteria.
To facilitate the search of the constrained diffeomorphism that results in a tile resembling the target shape, we develop a gradient-based optimization framework and introduce a novel loss function called \textit{Occupancy Loss}. The outline of our approach is depicted in Fig. \ref{fig:Methodology}. In the subsequent subsections, we delineate our framework and introduce the concept of Occupancy Loss in Section \ref{sec:occ_loss} and explain the optimization process in \ref{sec:optimization}. The construction of constrained diffeomorphisms tailored for Escherization tasks is elaborated in Section \ref{sec:constDiff}.

\begin{figure*}[]
  \centering
  \includegraphics[width=\linewidth]{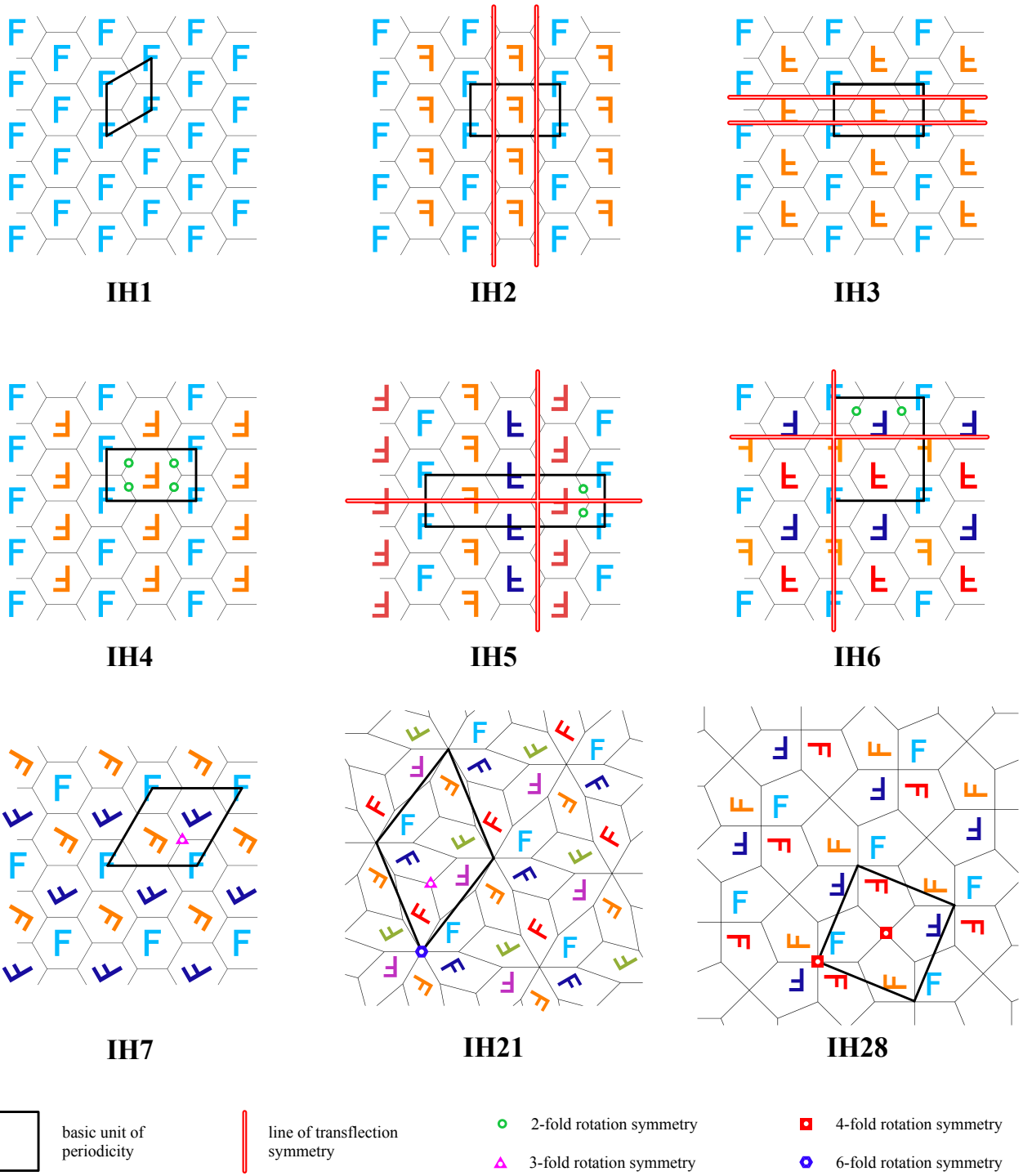}
  \caption{The tiling pattern for the nine isohedral classes and the symmetries in each are depicted. Each isohedral class is periodic, and their basic unit of periodicity is highlighted using a black-outlined quadrilateral. Further, there are other symmetries, such as k-fold rotations and transflections marked in each class. The red colored line indicates the line of transflection symmetry. Different types of rotation symmetries are marked using colored symbols. }
  \label{fig:IsohedralSymmetries}
\end{figure*}

\begin{table*}[]
\centering
\resizebox{\linewidth}{!}
{
\begin{tabular}{|c|c|c|c|c|}
\hline
 & Template Tile & Translational unit & Symmetries & Symmetrized Vector field \\ \hline
IH1 & 
\begin{minipage}{0.1\linewidth}
    \includegraphics[width=\linewidth]{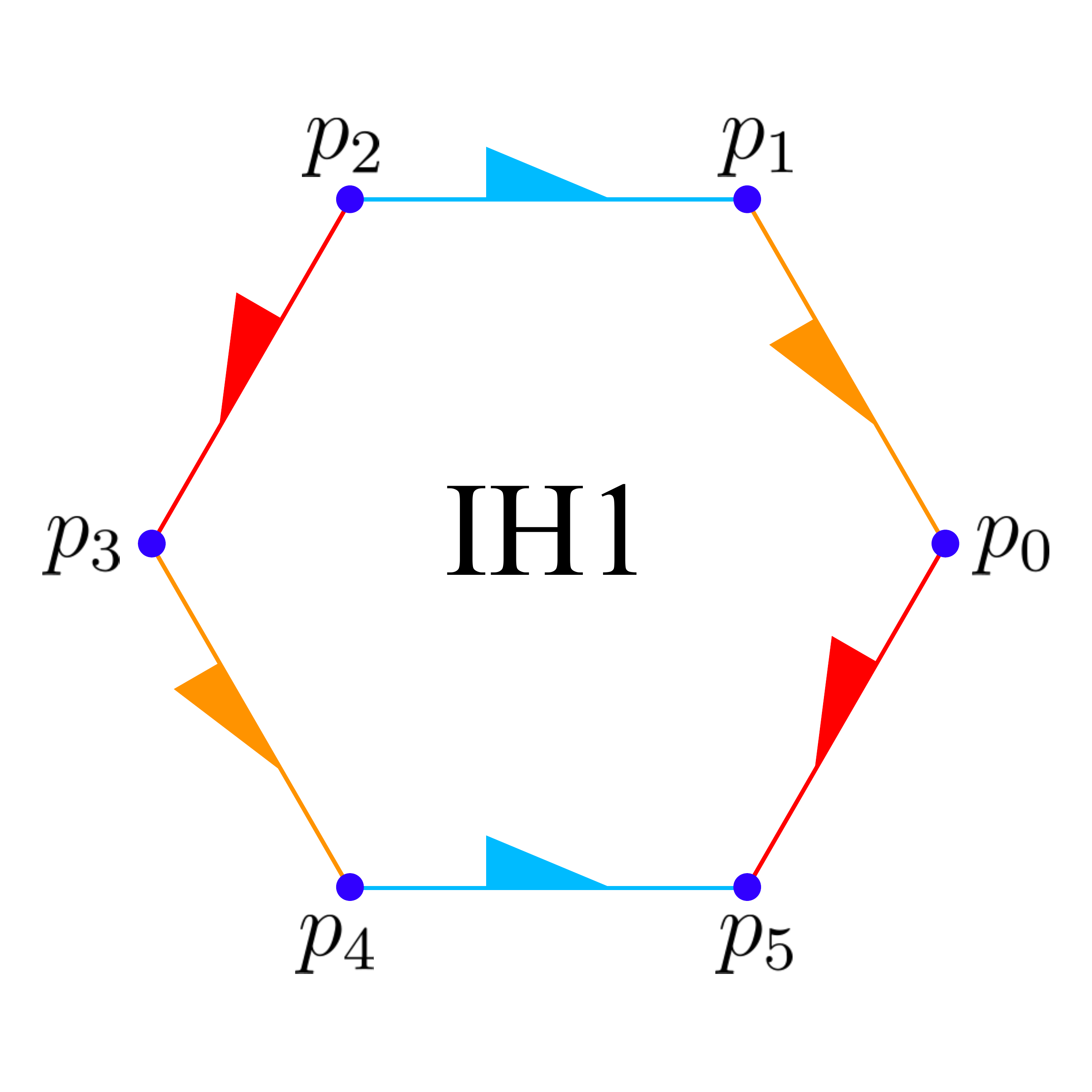} 
\end{minipage} &
\begin{tabular}[c]{@{}c@{}}$B = [b1,b2]$\\ $b1 = p_0 + p_1$\\ $b2 = p_1 + p_2$\end{tabular} & 
$B$-periodic & 
$V(z) = \mathcal{P}_B(F) \circ A^{-1}(z)$ \\ \hline

IH2 & 
\begin{minipage}{0.1\linewidth}
    \includegraphics[width=\linewidth]{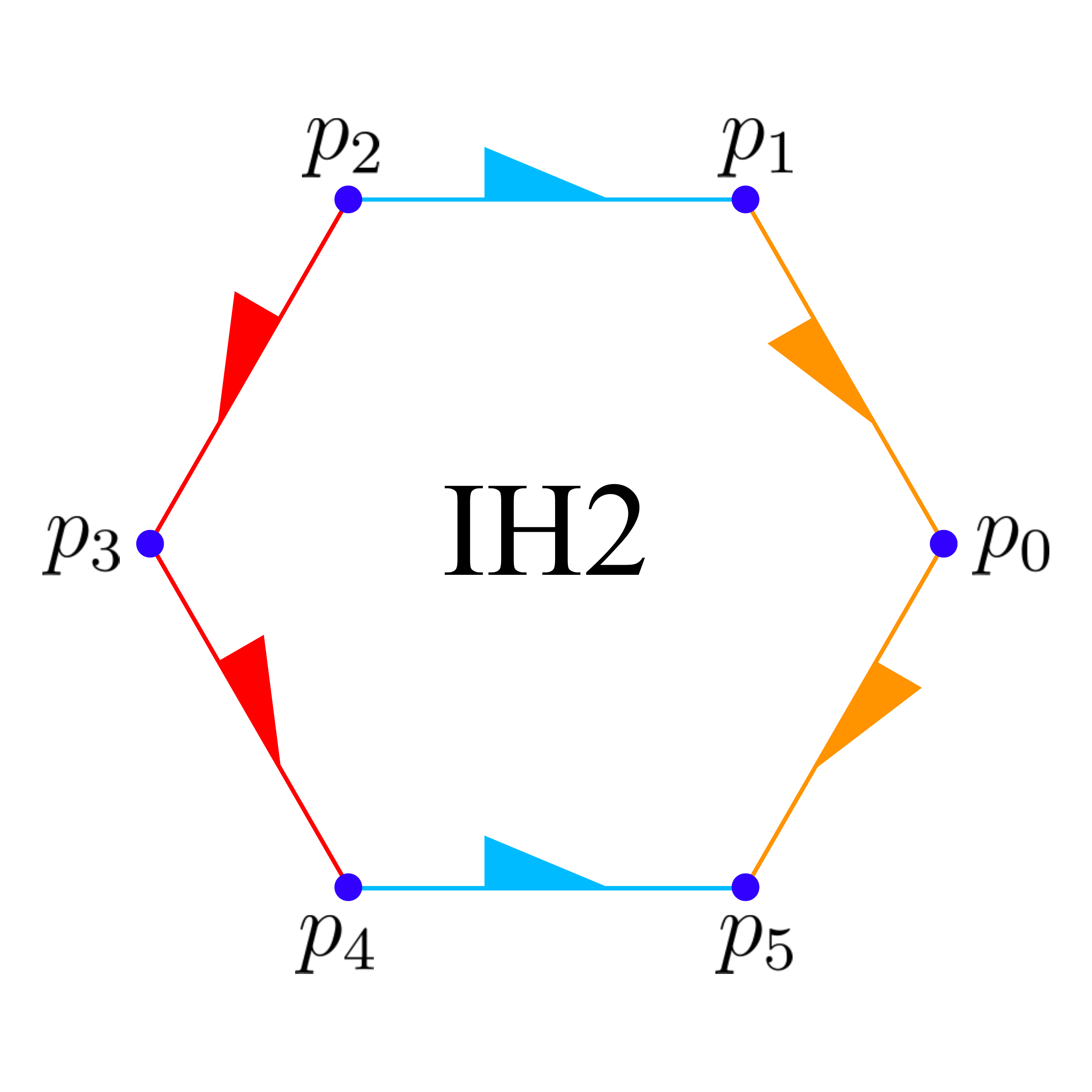} 
\end{minipage} &
\begin{tabular}[c]{@{}c@{}}$B = [b1,b2]$\\ $b1 = 3 p_0$\\ $b2 = p_1 + p_2$\end{tabular} & 
\begin{tabular}[c]{@{}c@{}}$B$-periodic\\ $(\ell_1,\mu_1)$-transflection\\ $(\ell_2,\mu_2)$-transflection\\ $\ell_1 \equiv {\frac{p_0 + p_1}{2}} + \lambda (\widehat{\frac{p_1 + p_2}{2}})$\\ $\ell_2 \equiv {\frac{5p_0 + p_2}{2}} + \lambda (\widehat{\frac{p_1 + p_2}{2}})$\\ $\mu_1 = \mu_2 = \Big|\Big|\frac{p_1 + p_2}{2}\Big|\Big|$\end{tabular} & 
$V(z) = \mathcal{G}_{\ell_1}^{\mu_1}(\mathcal{G}_{\ell_2}^{\mu_2}(\mathcal{P}_B(F))) \circ A^{-1}(z)$ \\ \hline
% $V(z) = \sum_{k_1=0}^{1}\sum_{k_2=0}^{1}M_{\ell_1^{o}}^{k_1}\cdot M_{\ell_2^{o}}^{k_2}\cdot B\cdot F\cdot B^{-1}\cdot {T_{\ell_2}^{\mu_2}}^{k_2}\cdot M_{\ell_2}^{k_2}\cdot {T_{\ell_1}^{\mu_1}}^{k_1}\cdot M_{\ell_1}^{k_1}\cdot A^{-1}(z)$ \\ \hline

IH3 & 
\begin{minipage}{0.1\linewidth}
    \includegraphics[width=\linewidth]{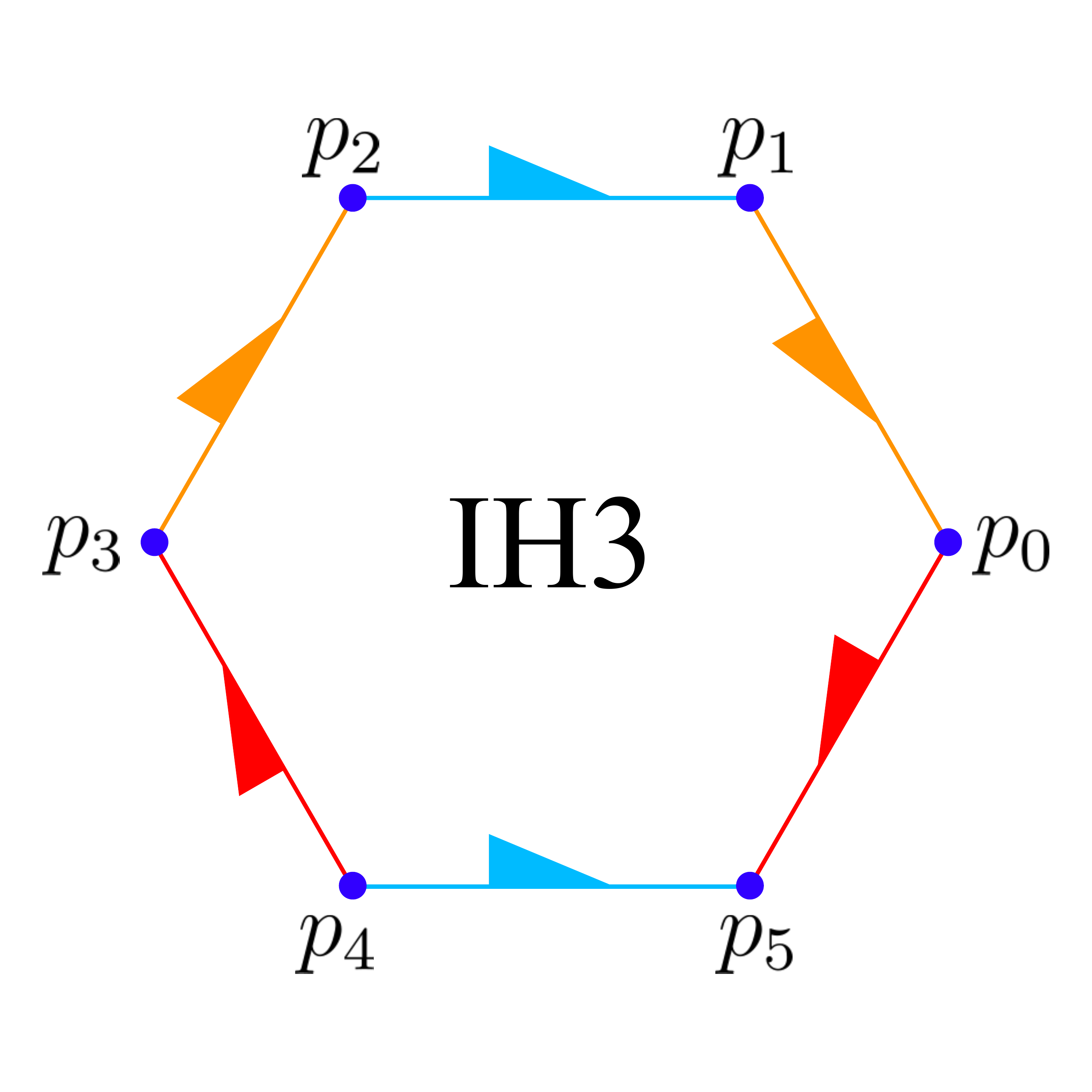} 
\end{minipage} &
\begin{tabular}[c]{@{}c@{}}$B = [b1,b2]$\\ $b1 = 3 p_0$\\ $b2 = p_1 + p_2$\end{tabular} & 
\begin{tabular}[c]{@{}c@{}}$B$-periodic\\ $(\ell_1,\mu_1)$-transflection\\ $(\ell_2,\mu_2)$-transflection\\ $\ell_1 \equiv {\frac{p_0 + p_1}{2}} + \lambda (\widehat{p_0})$\\ $\ell_2 \equiv {\frac{3}{4}(p_1 + p_2)} + \lambda (\widehat{p_0})$\\ $\mu_1 = \mu_2 = \Big|\Big|\frac{3 p_0}{2}\Big|\Big|$\end{tabular} &  
$V(z) = \mathcal{G}_{\ell_1}^{\mu_1}(\mathcal{G}_{\ell_2}^{\mu_2}(\mathcal{P}_B(F))) \circ A^{-1}(z)$ \\ \hline

IH4 & 
\begin{minipage}{0.1\linewidth}
    \includegraphics[width=\linewidth]{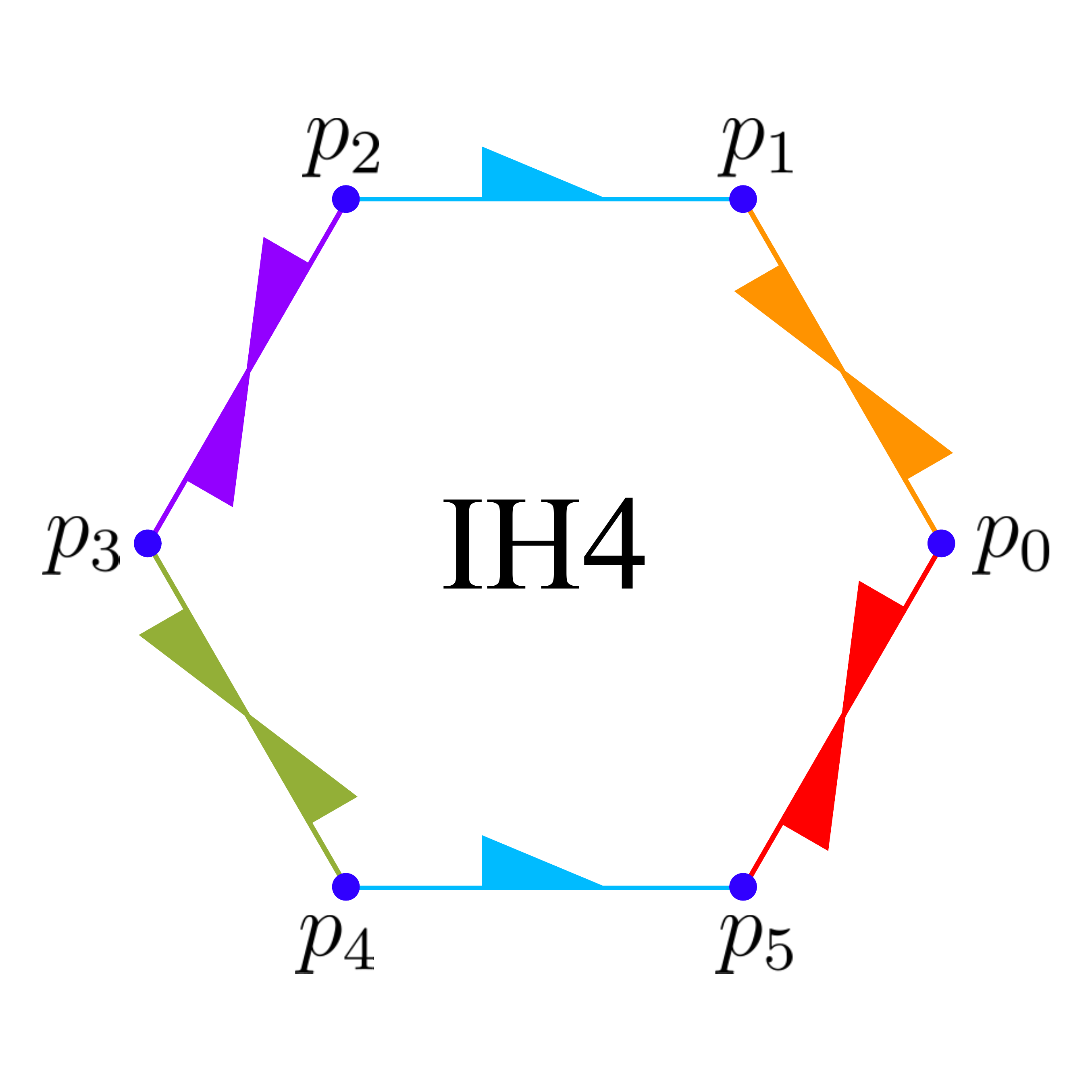} 
\end{minipage} &
\begin{tabular}[c]{@{}c@{}}$B = [b1,b2]$\\ $b1 = 3 p_0$\\ $b2 = p_1 + p_2$\end{tabular} & 
\begin{tabular}[c]{@{}c@{}}$B$-periodic\\ $(c_1,2)$-rotation, with $c_1 = \frac{3}{2}(p_0 + p_1)$\\ $(c_2,2)$-rotation with $c_2 = \frac{p_0 + 2 p_1 + p_2}{2}$\\ $(c_3,2)$-rotation with $c_3 = \frac{p_0 + p_1}{2}$\\ $(c_4,2)$-rotation with $c_4 = \frac{3 p_0 + 2 p_1 - p_2}{2}$\end{tabular} &  
$V(z) = \mathcal{R}_{c_1}^{2}(\mathcal{R}_{c_2}^{2}(\mathcal{R}_{c_3}^{2}(\mathcal{R}_{c_4}^{2}(\mathcal{P}_B(F))))) \circ A^{-1}(z)$ \\ \hline

IH5 & 
\begin{minipage}{0.1\linewidth}
    \includegraphics[width=\linewidth]{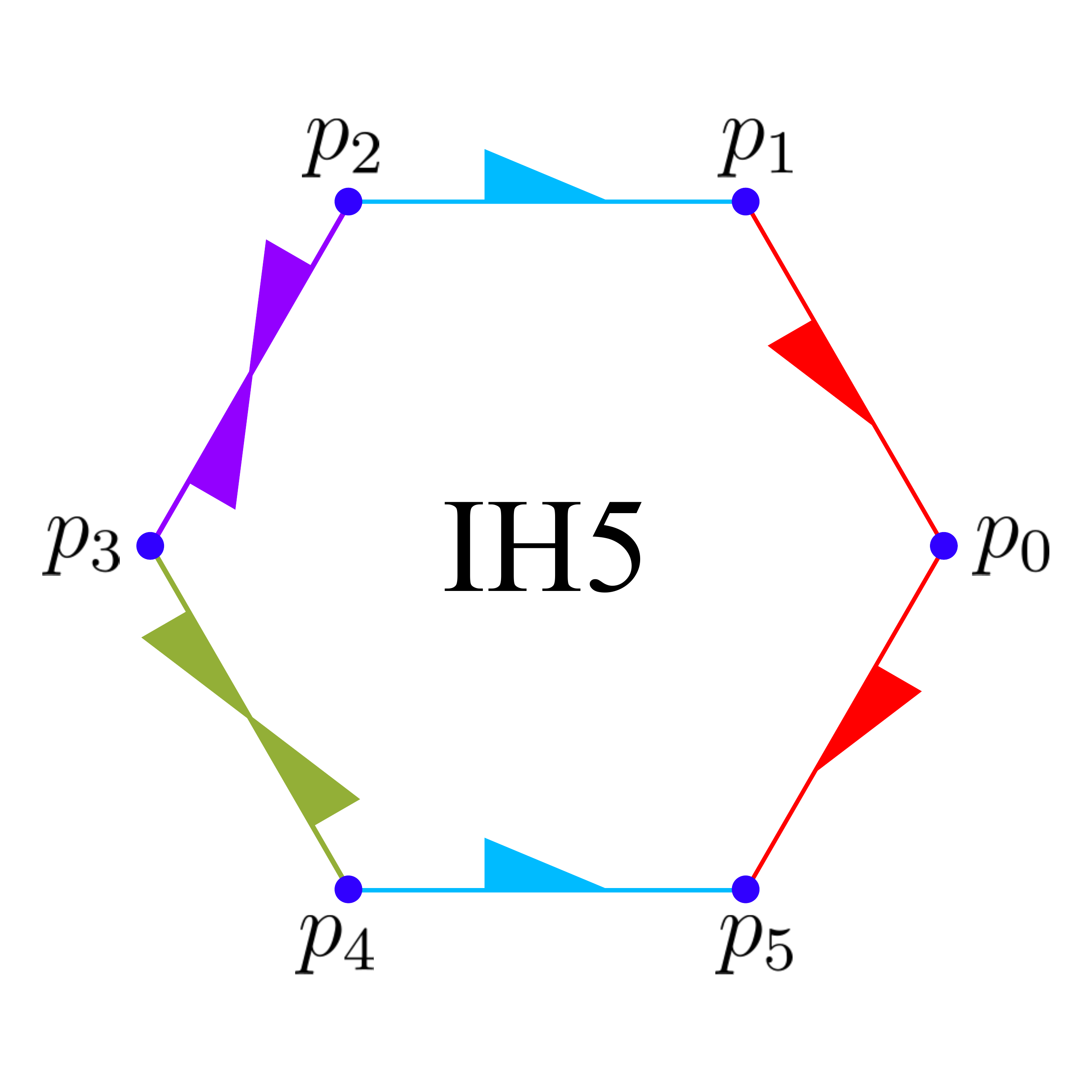} 
\end{minipage} &
\begin{tabular}[c]{@{}c@{}}$B = [b1,b2]$\\ $b1 = 6 p_0$\\ $b2 = p_1 + p_2$\end{tabular} & 
\begin{tabular}[c]{@{}c@{}}$B$-periodic\\ $(c_1,2)$-rotation, with $c_1 = \frac{11 p_0 + p_2}{2}$\\ $(c_2,2)$-rotation with $c_2 = \frac{11p_0 + p_1 + 2 p_2}{2}$\\ $(\ell_1,\mu_1)$-transflection\\ $(\ell_2,\mu_2)$-transflection\\ $\ell_1 \equiv {p_1} + \lambda (\widehat{p_0})$\\ $\ell_2 \equiv {\frac{7 p_0 + p_1}{2}} + \lambda (\widehat{p_1 + p_2})$\\ $\mu_1 = \Big|\Big|3 p_0\Big|\Big|$, $\mu_2 = \Big|\Big|\frac{p_1 + p_2}{2}\Big|\Big|$\end{tabular} &  
$V(z) = \mathcal{R}_{c_1}^{2}(\mathcal{R}_{c_2}^{2}(\mathcal{G}_{\ell_1}^{\mu_1}(\mathcal{G}_{\ell_2}^{\mu_2}(\mathcal{P}_B(F))))) \circ A^{-1}(z)$ \\ \hline

IH6 & 
\begin{minipage}{0.1\linewidth}
    \includegraphics[width=\linewidth]{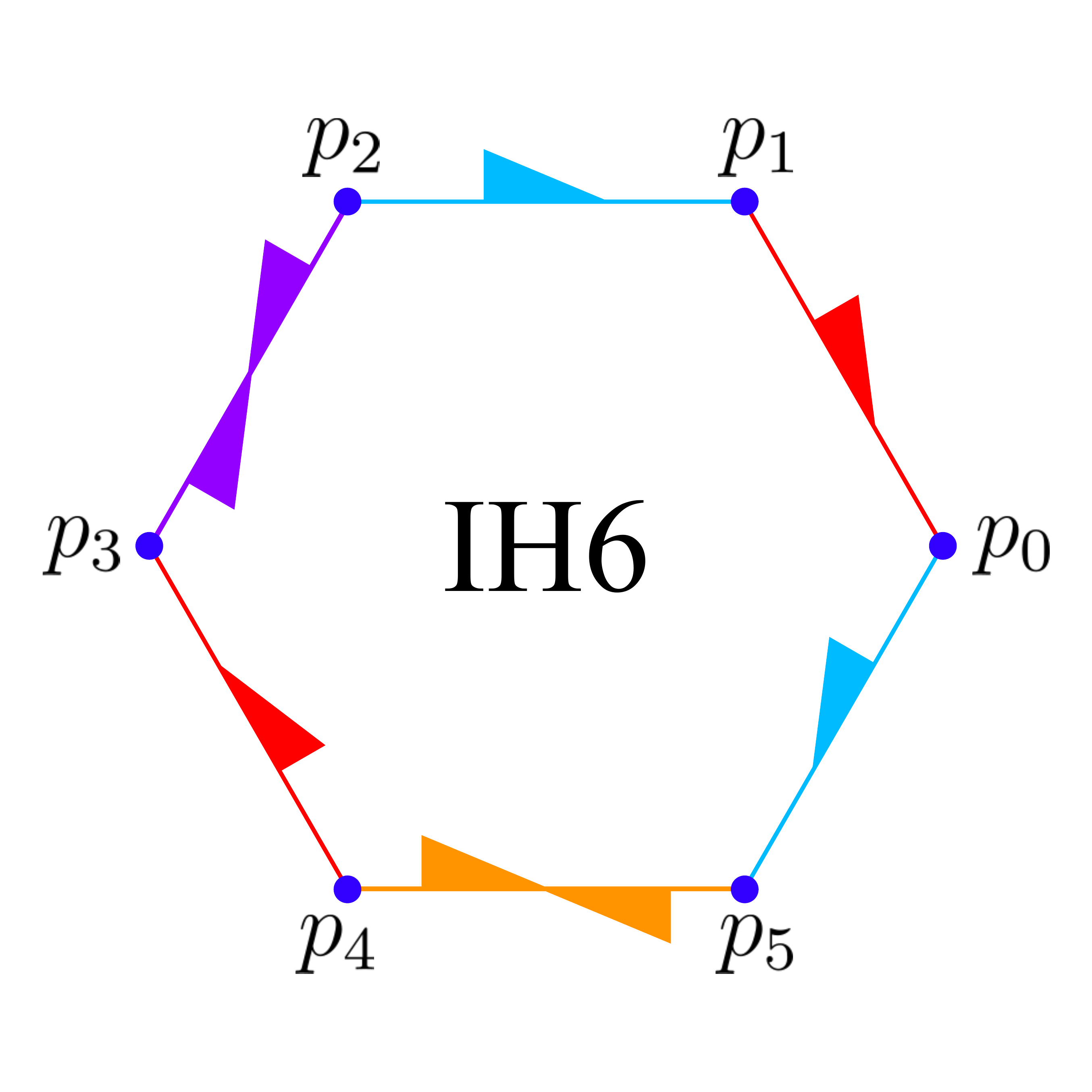} 
\end{minipage} &
\begin{tabular}[c]{@{}c@{}}$B = [b1,b2]$\\ $b1 = 3 p_0$\\ $b2 = 2 (p_1 + p_2)$\end{tabular} &  
\begin{tabular}[c]{@{}c@{}}$B$-periodic\\ $(c_1,2)$-rotation, with $c_1 = \frac{p_0 + 4 p_1 + 3 p_2}{2}$\\ $(c_2,2)$-rotation with $c_2 = \frac{5 p_0 + 3 p_1 + 4 p_2}{2}$\\ $(\ell_1,\mu_1)$-transflection\\ $(\ell_2,\mu_2)$-transflection\\ $\ell_1 \equiv \lambda (\widehat{p_1 + p_2})$\\ $\ell_2 \equiv {\frac{5}{4}(p_1 + p_2)} + \lambda (\widehat{p_0})$\\ $\mu_1 = \Big|\Big|p_1 + p_2\Big|\Big|$, $\mu_2 = \Big|\Big|\frac{3}{2}p_0\Big|\Big|$\end{tabular} &  
$V(z) = \mathcal{R}_{c_1}^{2}(\mathcal{R}_{c_2}^{2}(\mathcal{G}_{\ell_1}^{\mu_1}(\mathcal{G}_{\ell_2}^{\mu_2}(\mathcal{P}_B(F))))) \circ A^{-1}(z)$ \\ \hline

IH7 & 
\begin{minipage}{0.1\linewidth}
    \includegraphics[width=\linewidth]{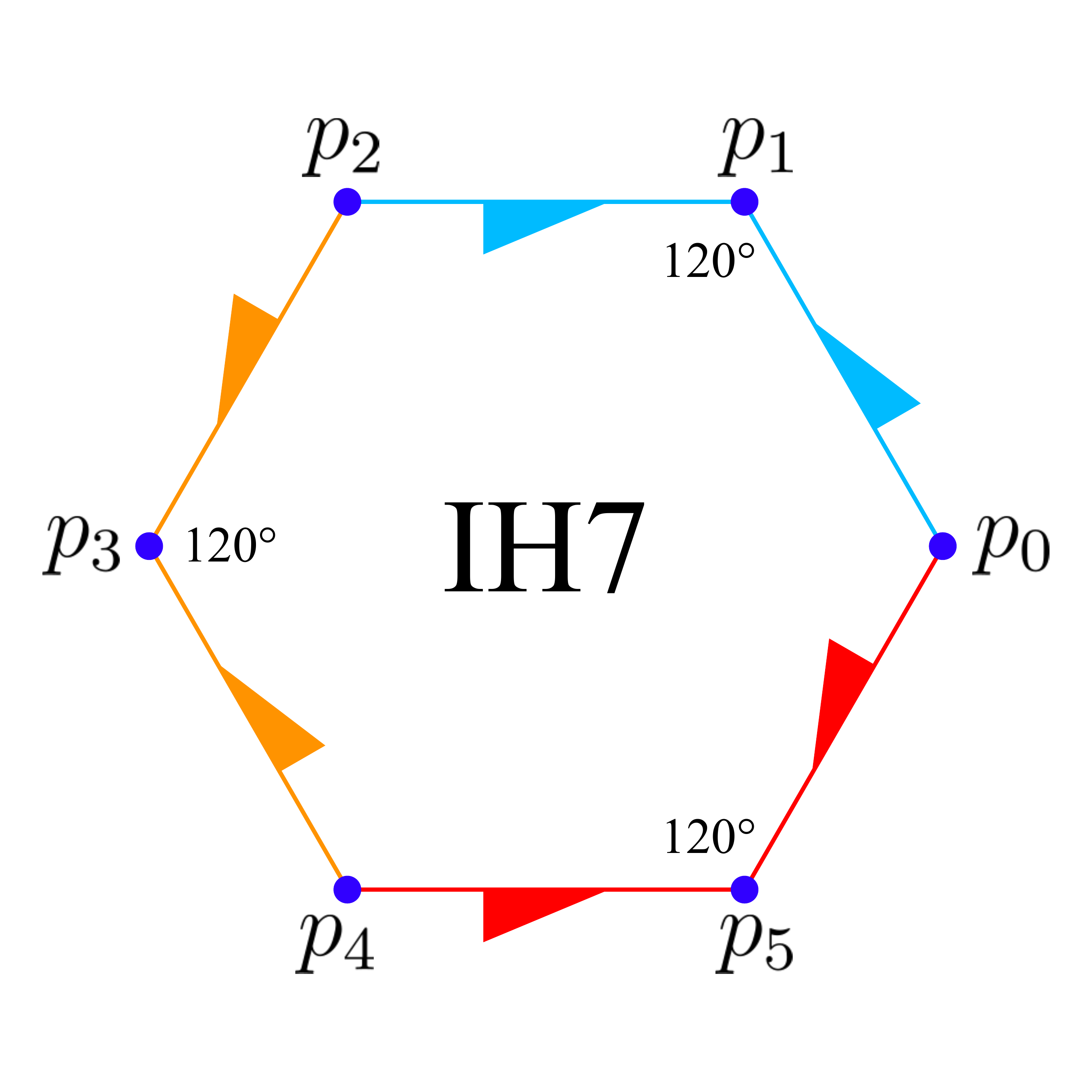} 
\end{minipage} &
\begin{tabular}[c]{@{}c@{}}$B = [b1,b2]$\\ $b1 = 3 p_0$\\ $b2 = 3 p_1$\end{tabular} &  
\begin{tabular}[c]{@{}c@{}}$B$-periodic\\ $(c,3)$-rotation, with $c = 2 p_0 + p_1$\end{tabular} &  
$V(z) = \mathcal{R}_{c}^{3}(\mathcal{P}_B(F)) \circ A^{-1}(z)$ \\ \hline

IH21 & 
\begin{minipage}{0.1\linewidth}
    \includegraphics[width=\linewidth]{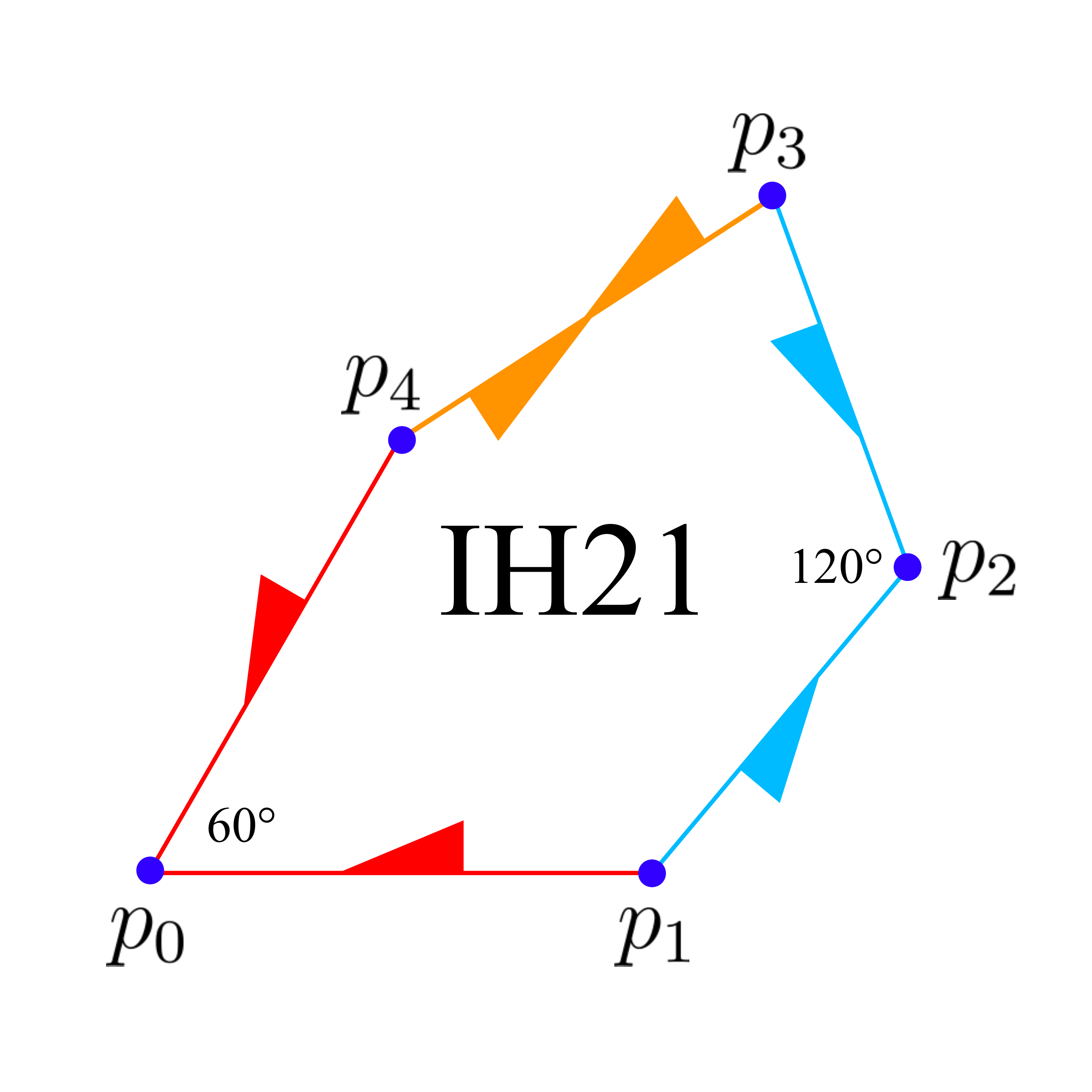} 
\end{minipage} &
\begin{tabular}[c]{@{}c@{}}$B = [b1,b2]$\\ $b1 = p_3 + p_4$\\ $b2 =\begin{bmatrix}\frac{1}{2} & -\frac{\sqrt{3}}{2}\\ \frac{\sqrt{3}}{2} & \frac{1}{2}\end{bmatrix} (p_3 + p_4)$\end{tabular} &  
\begin{tabular}[c]{@{}c@{}}$B$-periodic\\ $(c_1,3)$-rotation, with $c_1 = p_2$\\ $(c_2,6)$-rotation, with $c_2 = p_0$ \end{tabular}  &  
$V(z) = \mathcal{R}_{c_1}^{3}(\mathcal{R}_{c_2}^{6}(\mathcal{P}_B(F))) \circ A^{-1}(z)$ \\ \hline

IH28 & 
\begin{minipage}{0.1\linewidth}
    \includegraphics[width=\linewidth]{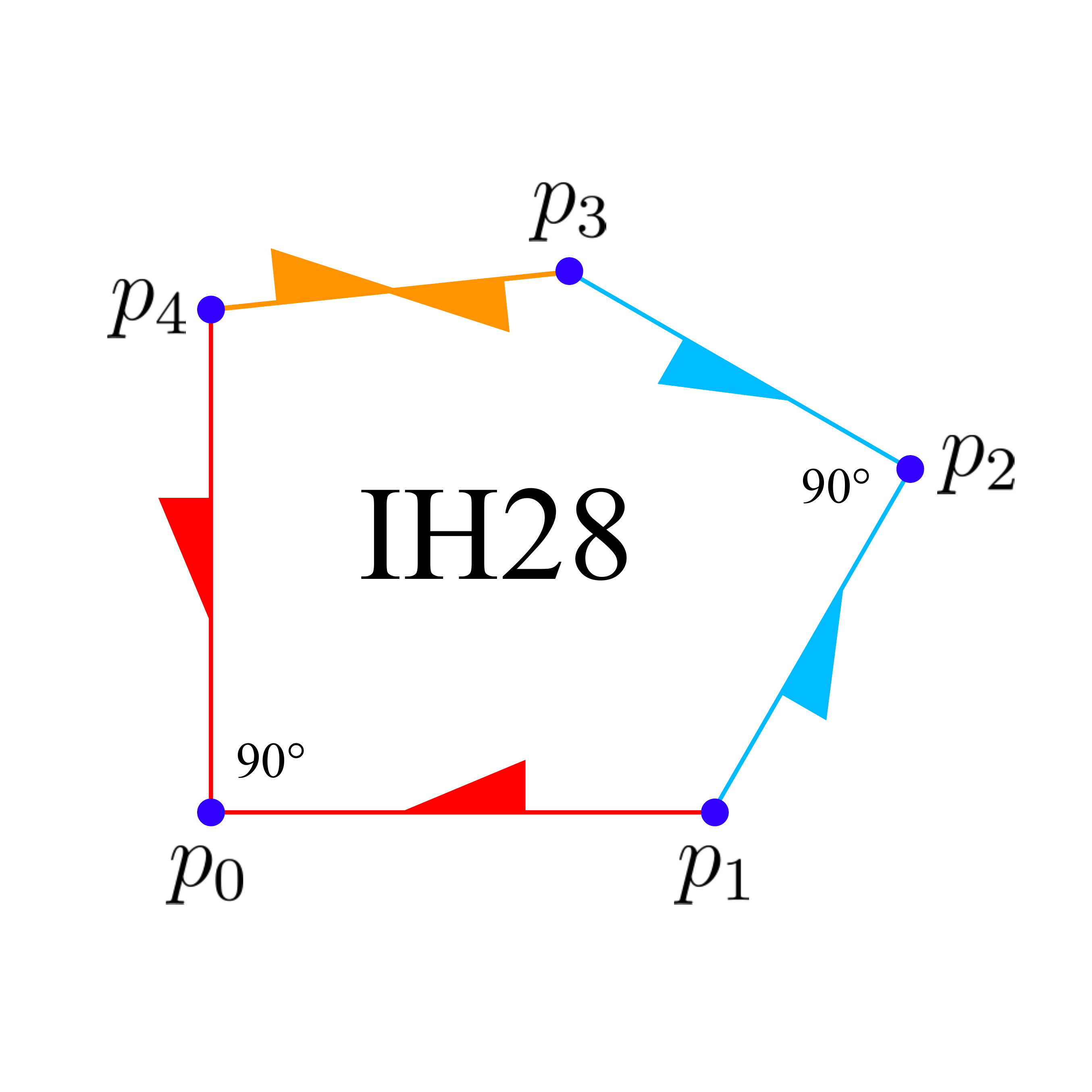} 
\end{minipage} &
\begin{tabular}[c]{@{}c@{}}$B = [b1,b2]$\\ $b1 = 2 p_2$\\ $b2 = p_3 + p_4$\end{tabular} & 
\begin{tabular}[c]{@{}c@{}}$B$-periodic\\ $(c_1,4)$-rotation, with $c_1 = p_2$\\ $(c_2,4)$-rotation, with $c_2 = p_0$ \end{tabular}  &  
$V(z) = \mathcal{R}_{c_1}^{4}(\mathcal{R}_{c_2}^{4}(\mathcal{P}_B(F))) \circ A^{-1}(z)$ \\ \hline
\end{tabular}
}
\caption{The Symmetrization for the nine isohedral classes are displayed. The first column depicts the identification constraint in each isohedral class. The second column conveys the basic translation unit we choose for our method. The third column describes in detail all the arising symmetries. The last column depicts the symmetrization equation to obtain a symmetric vector field. }
\label{tab:IH_symmetrization}
\end{table*}

\subsubsection{Occupancy Loss}
\label{sec:occ_loss}
% If we can learn a diffeomorphism that would map the interior points \( D_{\text{in}} \) of the template T with interior points \( Q_{\text{in}} \) of target shape S and exterior with exterior. Then this diffeomorphism would ensure that the template gets as close as possible to my target shape S. To train the Neural ODE for this shape transformation, we use the idea of invertibility. Specifically, we use the inverse of Neural ODE represented by eq: 

% \begin{equation}
% D(q)=s\left(t_0\right)=s\left(t_1\right)+\int_{t_0}^{t_1} V(s(t)) d t
% \end{equation}

To achieve a shape deformation from a template \( T \) to a target shape \( S \), we aim to find a constrained diffeomorphism \( D \) that maps the interior points of \( T \) to the interior points of \( S \), and similarly maps the exterior points of \( T \) to the exterior points of \( S \). Let \(P_{\text{in}}(T)\) and \(P_{\text{out}}(T)\) denote the interior and the exterior of $T$. Similarly, let \(P_{\text{in}}(S)\) and \(P_{\text{out}}(S)\) denote the interior and the exterior of $S$. Then our requirements can be expressed as:

\begin{equation}
D(P_{\text{in}}(T)) =  P_{\text{in}}(S) \quad \text{and} \quad D(P_{\text{out}}(T)) =  P_{\text{out}}(S)
\label{eq:occ_explain}
\end{equation}

The objective is to minimize the distance between the transformed template \( D(T) \) and the target shape \( S \), thereby ensuring that \( T \) approximates \( S \) as closely as possible. However, as explained later, we use the diffeomorphism in the inverse mode for practical purposes. The inverse of the diffeomorphism is given by:

\begin{equation}
D^{-1}(q) = s(t_0) = s(t_1) + \int_{t_1}^{t_0} V(s(t)) \, dt
\end{equation}

Here, \( D^{-1}(q) \) represents the inverse of diffeomorphism, \( s(t) \) is the state of the system at time \( t \), and \( V(s(t)) \) is the vector field induced by the diffeomorphism. The initial condition here is given by $s(t_1) = q$.

% So essentially we sample points from the target distribution and then use the Inverse NODE for taking these points back in the domain of my template distribution where due to the easy shape of the template it's feasible to easily create the sdf of the template shape. Mathematically given a point q inside the target S after diffeomorphics D(q) should land inside the template T. The distance of the point in template domain from the template is calculated using Sign-Distance function. 

% \begin{equation}
%     \phi(\mathbf{x})= \begin{cases}\min _{\mathbf{y} \in T}\|\mathbf{y}-\mathbf{x}\|, & \text { if } \mathbf{x} \text { is outside } T \\ 0, & \text { if } \mathbf{x} \in T \\ -\min _{\mathbf{y} \in T}\|\mathbf{y}-\mathbf{x}\|, & \text { if } \mathbf{x} \text { is inside } T\end{cases}
% \end{equation}

% We then apply the sigmoid function to create it into a format where we can apply the BCE loss function. 

% To the final loss function looks like :

% L = BCE (sigmoid(sdf(D(q)))

In essence, we sample points from the target shape \( S \) and employ the constrained diffeomorphism to map these points back to the domain of the template shape \( T \). Due to the geometric simplicity of \( T \)(either a hexagon or a pentagon), it is computationally efficient to compute its signed distance function (SDF). This is the reason for using the diffeomorphism in the inverse mode.
Let $q$ be a point in the domain of $S$. Let $o(q)$ denote the occupancy of $q$ in $S$. That is, $o(q)=1$ if $q$ is inside $S$ and $o(q)=0$ if $q$ is outside $S$. The inverse diffeomorphic image of $q$ be denoted as \( D^{-1}(q) \). The distance of this point from the boundary of \( T \) is evaluated using the Signed Distance Function (\( \phi \)) defined as

\begin{equation}
    \phi(\mathbf{x}) = \begin{cases}
    -\min_{\mathbf{y} \in \partial T} \|\mathbf{y} - \mathbf{x}\|, & \text{if } \mathbf{x} \text{ is outside } T, \\
    0, & \text{if } \mathbf{x} \in \partial T, \\
    \min_{\mathbf{y} \in \partial T} \|\mathbf{y} - \mathbf{x}\|, & \text{if } \mathbf{x} \text{ is inside } T.
    \end{cases}
\end{equation}
Here, $\partial T$ denotes the boundary of $T$.
Subsequently, we apply the sigmoid function \( \sigma(x) \) to the SDF values to obtain the soft occupancy values. The final loss function \( L \) is then given by

\begin{equation}
    L = \frac{1}{|Q|}\sum_{q \in Q}\text{BCE} \left( \sigma \left(\tau \cdot \phi(D^{-1}(q)) \right) , o(q) \right)
\label{eq:loss_func}
\end{equation}

where BCE denotes the Binary Cross-Entropy Loss, $\tau$ is the hardness factor, and $Q$ denotes a set of points sampled from the domain of $S$. This loss function tries to penalize the sample points that violate Eq. \ref{eq:occ_explain}. Thus, it tries to align $D(P_{\text{in}}(T))$ with $P_{\text{in}}(S)$ and $D(P_{\text{out}}(T))$ with $P_{\text{out}}(S)$, trying to ensure that the tile shape arising out of deformation resembles the target.

% In essence, the loss function is designed to ensure the following conditions:

% \begin{align*}
% L(Q_{\text{in}}, P_{\text{in}}) &\approx 0, & \text{when both points are interior} \\
% L(Q_{\text{out}}, P_{\text{out}}) &\approx 0, & \text{when both points are exterior} \\
% L(Q_{\text{out}}, P_{\text{in}}) &\approx 1, & \text{when \( Q \) is exterior but \( P \) is interior} \\
% L(Q_{\text{in}}, P_{\text{out}}) &\approx 1, & \text{when \( Q \) is interior but \( P \) is exterior}
% \end{align*}

% This loss function serves as an iterative optimization criterion for the Neural ODE, ensuring that \( Q_{\text{in}} \) aligns within \( P_{\text{in}} \) and \( Q_{\text{out}} \) falls outside \( P_{\text{out}} \).

% \subsubsection{Optimization process} We applied two fold optimization in this case, in first step we allow the template to undergo only Affine transformation, i.e. rotation and translation so as to speed up the convergence, as it trys to match the shape magnitude of S. Once the occupancy loss reduces lower than a certain threshold then  we start the Constrained flow based optimization process for letting the template shape learn the target shape.  

\subsubsection{Optimization Process}
\label{sec:optimization}

Given a parametric family of constrained diffeomorphisms $D_{\theta}$ for $\theta \in \Theta$, our objective is to find the parameter $\theta$ so that the diffeomorphism $D_{\theta}$ minimizes the loss function defined in Eq. \ref{eq:loss_func}.
Our optimization method consists of two phases.

\begin{enumerate}
    \item \textbf{Affine Transformation Phase:} Initially, we restrict the template to undergo only affine transformations, specifically rotations, translations, and uniform scaling. This phase aims to quickly align the template with the target shape \( S \) in terms of orientation, location, and scale. We find the optimal affine transformation $A$ that aligns the template tile in the best possible way.
    
    % The optimization continues until the Occupancy Loss falls below a predefined threshold \( \tau \).
    
    % \begin{equation}
    %     \text{Minimize } L(T, S) \text{ subject to } T = \text{Affine}(T) \text{ until } L(T, S) < \tau
    % \end{equation}
    
    \item \textbf{Constrained Flow Phase:} 
    Here, we use a composition of the affine transformation and the constrained diffeomorphism. In this phase, the template shape is allowed to deform freely within the constraints defined by its IH class symmetry, aiming to find a tile that resembles the target shape \( S \).

\end{enumerate}

\subsubsection{Constrained Diffeomorphisms for Isohedral Classes}

\label{sec:constDiff}

While deforming the template tile of a particular IH class, we need to ensure that the deformed tile belongs to that IH class. For this, we need to ensure that the identification constraints of the IH class are encoded in the constrained diffeomorphism. In our case, diffeomorphism comes from integrating over a vector field. The vector field needs to have particular symmetries arising from the identification constraints. To make the vector field symmetric, we use the process of symmetrization as mentioned in Sec. \ref{sec:Symmetrization}. The symmetries arising in each IH class are visually depicted in Fig. \ref{fig:IsohedralSymmetries}. The symmetries and the symmetrization equation are explained in detail in Tab. \ref{tab:IH_symmetrization}. Note that in the symmetrization equations $V$ is the symmetrized vector field, $\mathcal{P}_B$, $\mathcal{G}_{\ell}^{\mu}$, and $\mathcal{R}_{c}^{K}$ denote the symmetrization operators (defined in Sec. \ref{sec:Symmetrization}), $F$ is the periodic vector field on the canonical basis (defined in Eq. \ref{eq:F_function}), and $A$ is the affine transformation defined in Sec. \ref{sec:optimization}.

% In our \texttt{IH01} design, as illustrated in fig: \ref{fig:IsohedralSymmetries}, we employ a lattice constituted of hexagonal cells, each distinctly marked with the letter 'F'. At the heart of this design, a rhombus stands out, acting as the fundamental domain or the elemental block for our escherization process. This rhombus embodies the primary design and flow. With the employment of our constrained flow methods, affine transformations become pivotal, guiding the design within this rhombus to adapt and fit within the constraints of the surrounding hexagons. These transformations ensure the design's integrity and consistency as it repeats. Exploiting the principles of translational symmetry, the design encapsulated within the rhombus is then propagated seamlessly throughout the hexagonal grid, with canonical basis vectors orchestrating the orientation and positioning of these replications. The result is a consistent and intricate flow across the entire structure, where every hexagon in the lattice harmoniously mirrors the design from the rhombus. Through this methodology, rooted in the rhombic fundamental domain and bolstered by affine transformations, we guarantee impeccable escherization.

\begin{figure}[]
  \centering
  \includegraphics[width=0.95\linewidth]{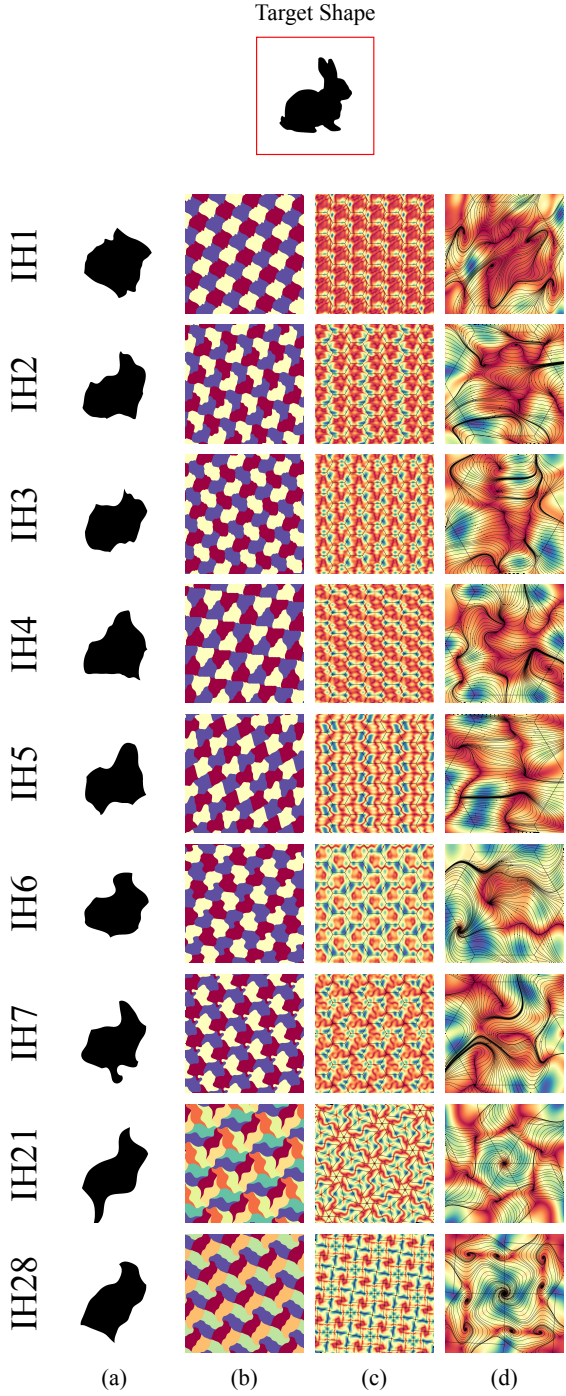}
  \caption{Given the target image of a rabbit, the results for Escherization using our approach are shown for the 9 IH classes. (a) The tile shape that satisfies the constraints of the IH class while trying to resemble the rabbit and (b) the corresponding tiling the tile shape admits are shown. The (c) magnitude of the vector flow field and (d) the vector flow lines are visualized. It can be seen that the vector flow field has the same symmetry that is inherent in the symmetry group of each IH class. These results are computed for $\Omega=5$.}
  \label{fig:NP0}
\end{figure}

\begin{figure*}[]
  \centering
  \includegraphics[height=0.9\textheight]{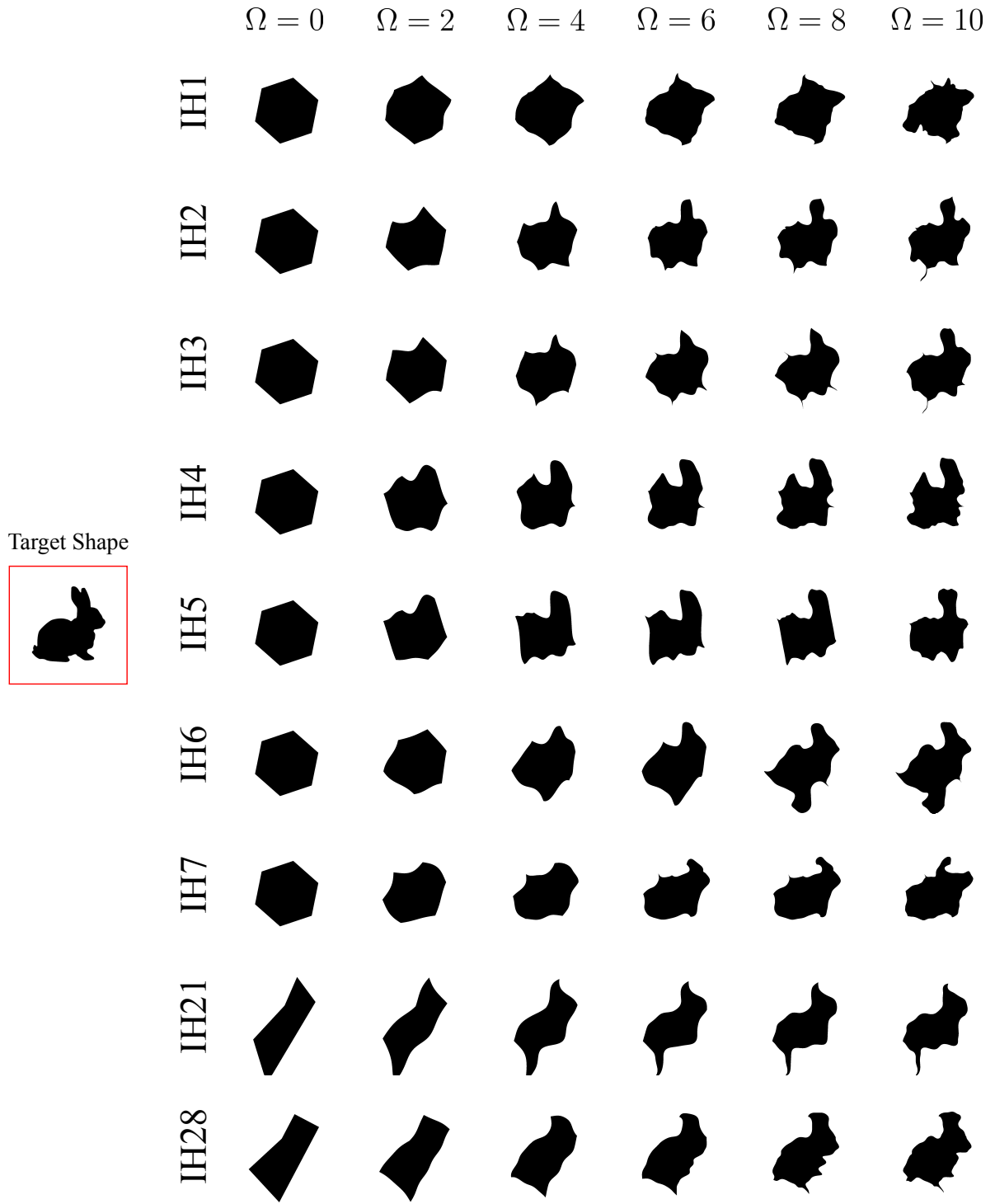}
  \caption{The effect of changing the maximum frequency $\Omega$ is visualized. For $\Omega=0$, there is no deformation in the shape of the tile as the diffeomorphism only has translational effects. On increasing the frequency, we observe that the template exhibits intricate deformations. We can see that the higher the allowed frequency, the more complex the tile shape.}
  \label{fig:FourierVariation}
\end{figure*}

\begin{figure*}[]
  \centering
  \includegraphics[width=\linewidth]{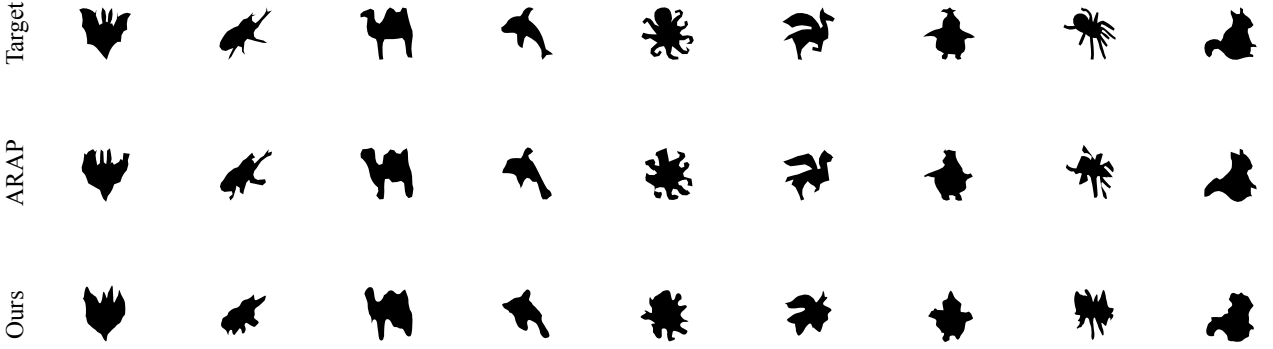}
  \caption{Comparison of our method with ARAP \cite{nagata2021escherization} for 9 different target shapes. The first row contains the nine target images. The second row depicts the resulting tile shape obtained using ARAP \cite{nagata2021escherization}, and the third row depicts the resulting tile shape obtained using our method.}
  \label{fig:EscherizationComparison}
\end{figure*}

\subsection{Results and Experiments}
The Escherization results were obtained on two distinct datasets. The first dataset replicates the one used in \cite{nagata2021escherization}, while the second is a custom dataset curated from \cite{Keesey2023}. Our motivation for creating a custom dataset stems from the polygonal limitations of the dataset in \cite{nagata2021escherization}. As our method is capable of generating general tile shapes, we can run it on a diverse dataset that includes non-polygonal targets.

Performance comparisons with \cite{nagata2021escherization} are presented in Figure \ref{fig:EscherizationComparison} and Table \ref{table:comparison_metrics}. Our method demonstrates comparable performance and offers two key advantages. 
Firstly, our method supports general tile shapes without being confined to polygonal forms. Secondly, our method works on target shapes that are non-polygonal given in the form of silhouette images. The ARAP method \cite{nagata2021escherization} is specifically tailored for polygonal target shapes. Our method, despite being designed for general shapes, obtains competitive performance for polygonal targets. 

The second dataset, being non-polygonal, could only be evaluated using our method due to the limitations of other approaches only addressing polygonal shapes. One of the results is displayed in Figure \ref{fig:NP0}, while ten additional results are included in the supplementary material. Figure \ref{fig:NP0} showcases outputs for nine IH classes, their corresponding tiling patterns in \( \mathbb{R}^2 \), the constrained diffeomorphic vector flow field (indicating the incorporated symmetries for each IH class), and the vector flow lines.

We further conduct an ablation study to observe the effect of changing the maximum allowed frequency $\Omega$ as shown in Fig. \ref{fig:FourierVariation}. We can see that as the frequency is increased, the tile shapes become more complex for each IH class. For frequency $\Omega=0$, the tile shape is equivalent to the template tile as $\Omega=0$ only permits translational changes. Thus, our method successfully generates complex non-polygonal tile shapes, and the complexity increases as we increase the frequency.

% In our \texttt{IH01} design, as illustrated in fig: \ref{fig:IsohedralSymmetries}, we employ a lattice constituted of hexagonal cells, each distinctly marked with the letter 'F'. At the heart of this design, a rhombus stands out, acting as the fundamental domain or the elemental block for our escherization process. This rhombus embodies the primary design and flow. With the employment of our constrained flow methods, affine transformations become pivotal, guiding the design within this rhombus to adapt and fit within the constraints of the surrounding hexagons. Aiding in this optimization is our Binary Cross-Entropy (BCE) with sigmoid loss function, ensuring that the transformed points align precisely within the template domain. These transformations ensure the design's integrity and consistency as it repeats. Exploiting the principles of translational symmetry, the design encapsulated within the rhombus is then propagated seamlessly throughout the hexagonal grid, with canonical basis vectors orchestrating the orientation and positioning of these replications. The result is a consistent and intricate flow across the entire structure, where every hexagon in the lattice harmoniously mirrors the design from the rhombus. Through this methodology, rooted in the rhombic fundamental domain and bolstered by affine transformations and precise loss optimization, we guarantee impeccable escherization.

\begin{table}[!t]
\centering
\resizebox{0.45\textwidth}{!}{%
\begin{tabular}{c|c|c|c|c}
\toprule
\multicolumn{1}{c}{} & \multicolumn{2}{|c|}{IoU} & \multicolumn{2}{c}{Pixel-Accuracy} \\
\cmidrule(lr){2-5}
Image & Ours & ARAP \cite{nagata2021escherization} & Ours & ARAP \cite{nagata2021escherization} \\
\midrule
Bat & 0.859 & 0.869 & 0.880 & 0.895 \\
Beetle & 0.917 & 0.929 & 0.909 & 0.927 \\
Camel & 0.860 & 0.875 & 0.883 & 0.903 \\
Dolphin & 0.924 & 0.944 & 0.918 & 0.942 \\
Octopus & 0.809 & 0.818 & 0.834 & 0.849 \\
Pegasus & 0.872 & 0.868 & 0.880 & 0.882 \\
Penguin & 0.927 & 0.935 & 0.927 & 0.941 \\
Spider & 0.856 & 0.856 & 0.846 & 0.856 \\
Squirrel & 0.893 & 0.922 & 0.906 & 0.935 \\
\bottomrule
\end{tabular}%
}
\caption{Comparison of IoU and Pixel-Accuracy between Ours and ARAP \cite{nagata2021escherization}.}
\label{table:comparison_metrics}
\end{table}

\section{Density Estimation on Manifolds}
\label{sec:FlowOnManifolds}

%%%%%%%%%%%%%%%%%%%%%%%%%%%%%%%%%%%%%%%%%%%%%%%%
% Defining diffeomorphisms on euclidean space easy, challenging on non-euclidean manifolds
% In this work we define a way of designing diffeomorphisms on identification spaces
% Several shapes such as T,S,K,P can be obtained through identification topology as shown in Fig.
% Given a square with identification constraints, results in different topological spaces.
% Given a square patch with identiication constraints, if we make countable copies of the square and aseemble them based on their identification constraints, we get a tiling pattern as shown in the figure. The tiling pattrern is different for each cases S,T,K,P. These tilings patterns involve symmetry constraints such as periodicity, rotation and transflection. 
% >>>> Canonical Projection <<<<<
% In the tiling, Lets call the unit square $[0,1]^2$ as the cannonical square. If we start moving from a point on the canonical cell and reach a point in a different cell, since we know that the cell is a copy of the canonical cell, the new point also belongs to the original cell. The operation of finding the projection of this new point on the canonical cell, is termned here as cannonical projection.
%
%%%%%%%%%%%%%%%%%%%%%%%%%%%%%%%%%%%%%%%%%%%%%%%%

\label{sec:density-estimation}

% \section{Density Estimation}
Let \( x \) be a continuous random variable in \( \mathbb{R}^d \) with an unknown probability density function \( p(x) \). Given a sample data set \( X = \{x_1, x_2, \ldots, x_N\} \) consisting of \( N \) i.i.d. observations drawn from \( p \), the objective of density estimation is to construct a density function \( \hat{p} \) that closely approximates \( p \).

A prevalent metric for evaluating the quality of the approximation is the Kullback-Leibler (KL) divergence, denoted as \( L_{\text{KL}}(p || \hat{p}) \). Alternatively, one could employ other distance metrics \( d(p, \hat{p}) \) tailored to the problem. The aim is to find \( \hat{p} \) that minimizes this distance to \( p \) across the data \( X \).

% \begin{equation}
% \label{eq:KL_divergence}
% D_{KL}(p || \hat{p}) = \int_{\mathbb{R}^d} p(x) \log \left(\frac{p(x)}{\hat{p}(x)}\right) dx.
% \end{equation}

\subsection{Normalizing Flows} Normalizing flows offer a principled approach for density estimation by transforming a simple template distribution through a sequence of invertible and differentiable transformations to model a complex target distribution. Given a simple template distribution $q_0(x)$, such as a multivariate Gaussian, a normalizing flow applies a sequence of $T$ transformations $f_t: \mathbb{R}^d \rightarrow \mathbb{R}^d$, for $t = 1, \ldots, T$, to obtain the final transformed distribution $q_T(x)$ which approximates the target distribution. The transformations are chosen so that the estimated density $q_T(x)$ is easy to evaluate and sample. The density of $q_T(x)$ can be computed using the change of variables formula:

\begin{equation}
\label{eq:change_of_variables}
q_T(x) = q_0(f_1^{-1} \circ \ldots \circ f_T^{-1}(x)) \cdot \left| \det \left( \frac{\partial(f_1^{-1} \circ \ldots \circ f_T^{-1})}{\partial x} \right) \right|
\end{equation}

% where $\det$ denotes the determinant of the Jacobian matrix of the transformations. 

% \subsection{Limitation of Normalizing Flows:} 

While Normalizing Flows excels at constructing diffeomorphisms in Euclidean spaces like \( \mathbb{R}^2 \), extending the same to non-Euclidean manifolds is challenging. This is mainly due to the fact that there does not exist a smooth, invertible map for projecting the Normalizing Flows from the Euclidean spaces to the non-Euclidean spaces. In this work, we use Flow Symmetrization to overcome these issues and design Normalizing Flows on non-Euclidean spaces, specifically on identification spaces such as torus, sphere, Klein bottle, and projective plane.

\begin{table*}[]
\centering
\resizebox{\linewidth}{!}
{
\begin{tabular}{|c|c|c|c|c|}
\hline
 & Template Tile & Translational unit & Symmetries & Symmetrized Vector field \\ \hline
Torus & 
\begin{minipage}{0.1\linewidth}
    \includegraphics[width=\linewidth]{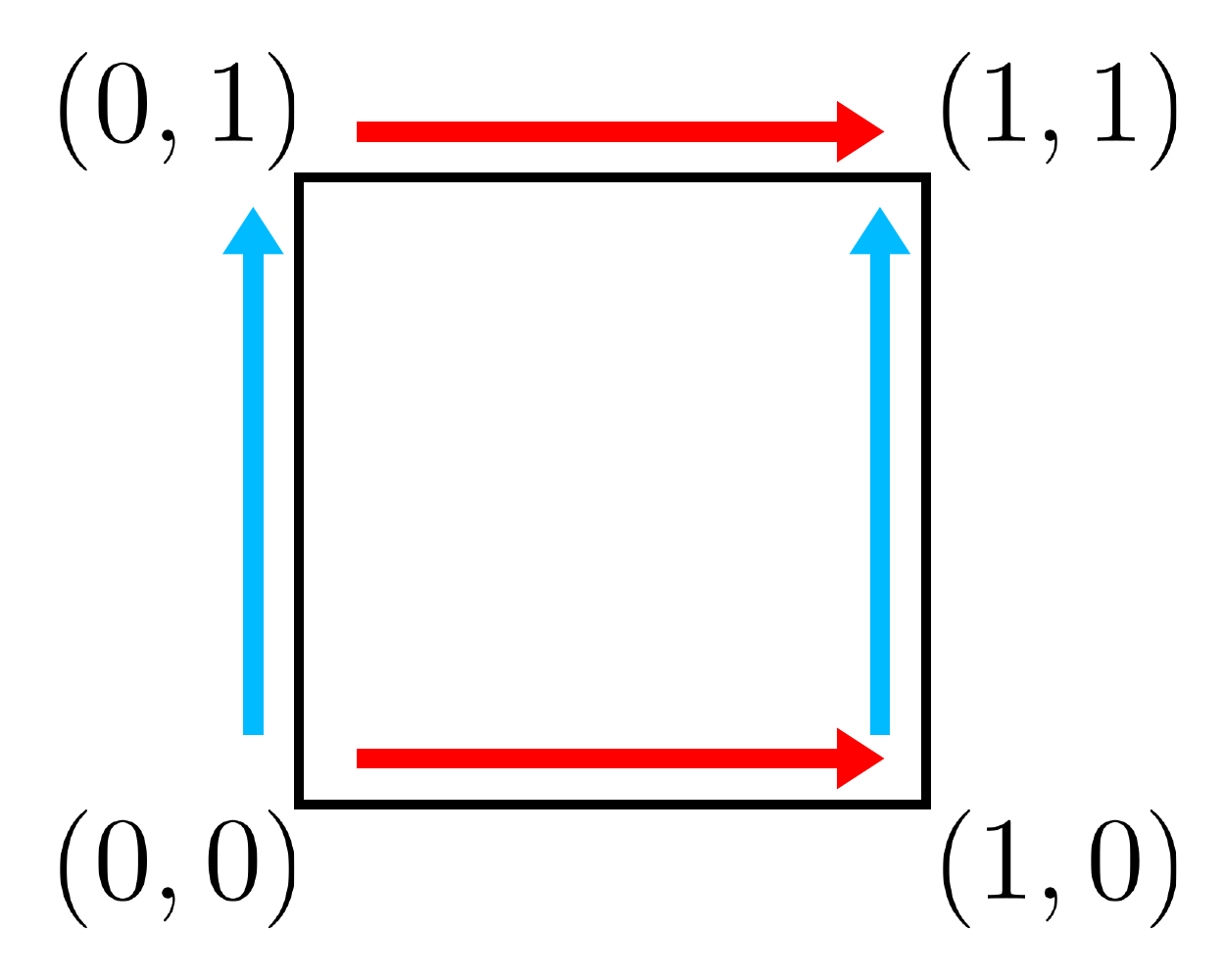} 
\end{minipage} &
\begin{tabular}[c]{@{}c@{}}$B = [b1,b2]$\\ $b1 = [1, 0]$\\ $b2 = [0, 1] $\end{tabular} & 
$B$-periodic & 
$V(z) = \mathcal{P}_B(F)(z)$ \\ \hline

Sphere & 
\begin{minipage}{0.1\linewidth}
    \includegraphics[width=\linewidth]{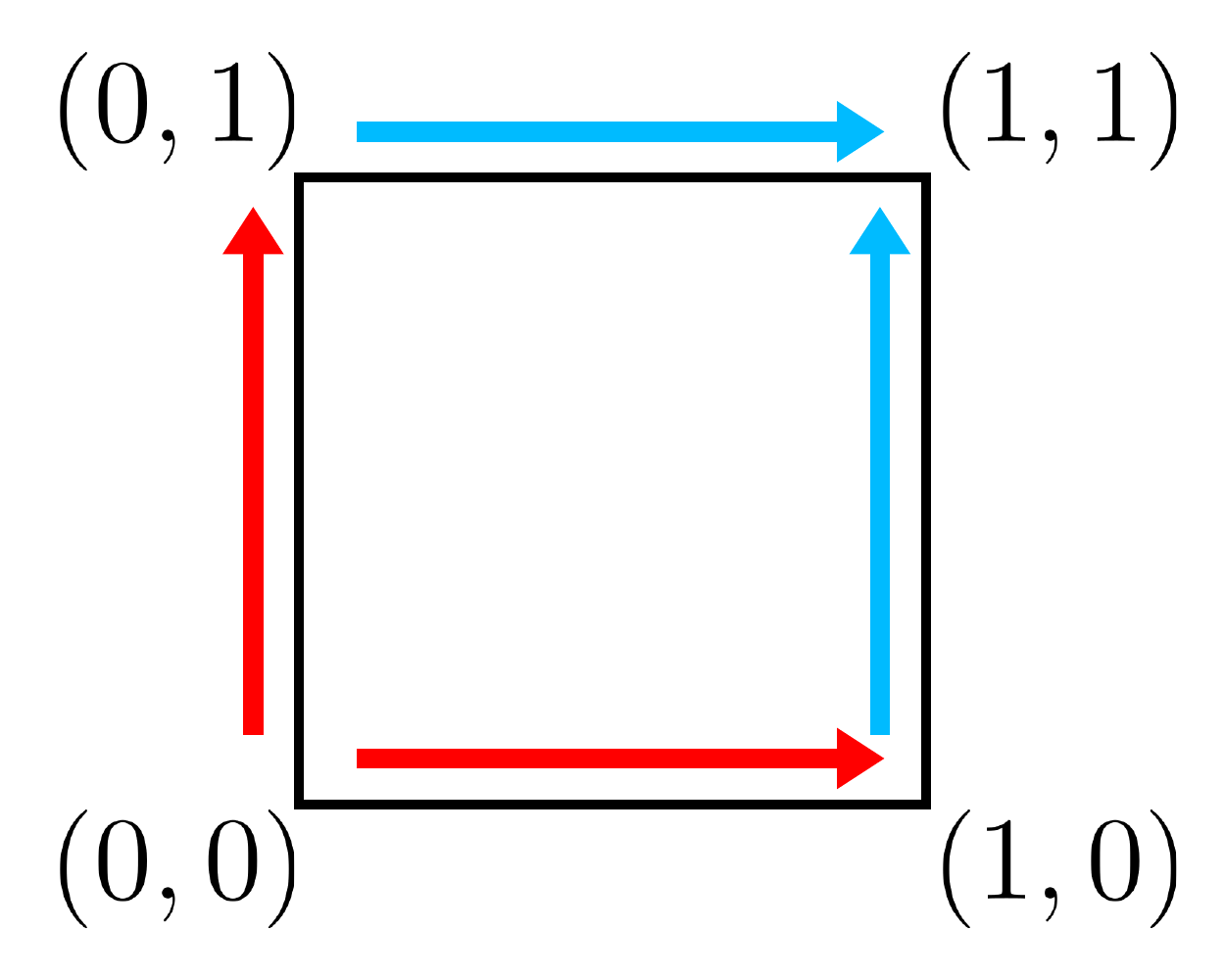} 
\end{minipage} &
\begin{tabular}[c]{@{}c@{}}$B = [b1,b2]$\\ $b1 = [2, 0]$\\ $b2 = [0, 2]$\end{tabular} & 
\begin{tabular}[c]{@{}c@{}}$B$-periodic\\ $(c,4)$-rotation, with $c = [1,1]$\ \end{tabular} & 
$V(z) = \mathcal{R}_{c}^{4}(\mathcal{P}_B(F))(z)$ \\ \hline
% $V(z) = \sum_{k_1=0}^{1}\sum_{k_2=0}^{1}M_{\ell_1^{o}}^{k_1}\cdot M_{\ell_2^{o}}^{k_2}\cdot B\cdot F\cdot B^{-1}\cdot {T_{\ell_2}^{\mu_2}}^{k_2}\cdot M_{\ell_2}^{k_2}\cdot {T_{\ell_1}^{\mu_1}}^{k_1}\cdot M_{\ell_1}^{k_1}\cdot A^{-1}(z)$ \\ \hline

\begin{tabular}[c]{@{}c@{}}Klein\\ Bottle\end{tabular}  & 
\begin{minipage}{0.1\linewidth}
    \includegraphics[width=\linewidth]{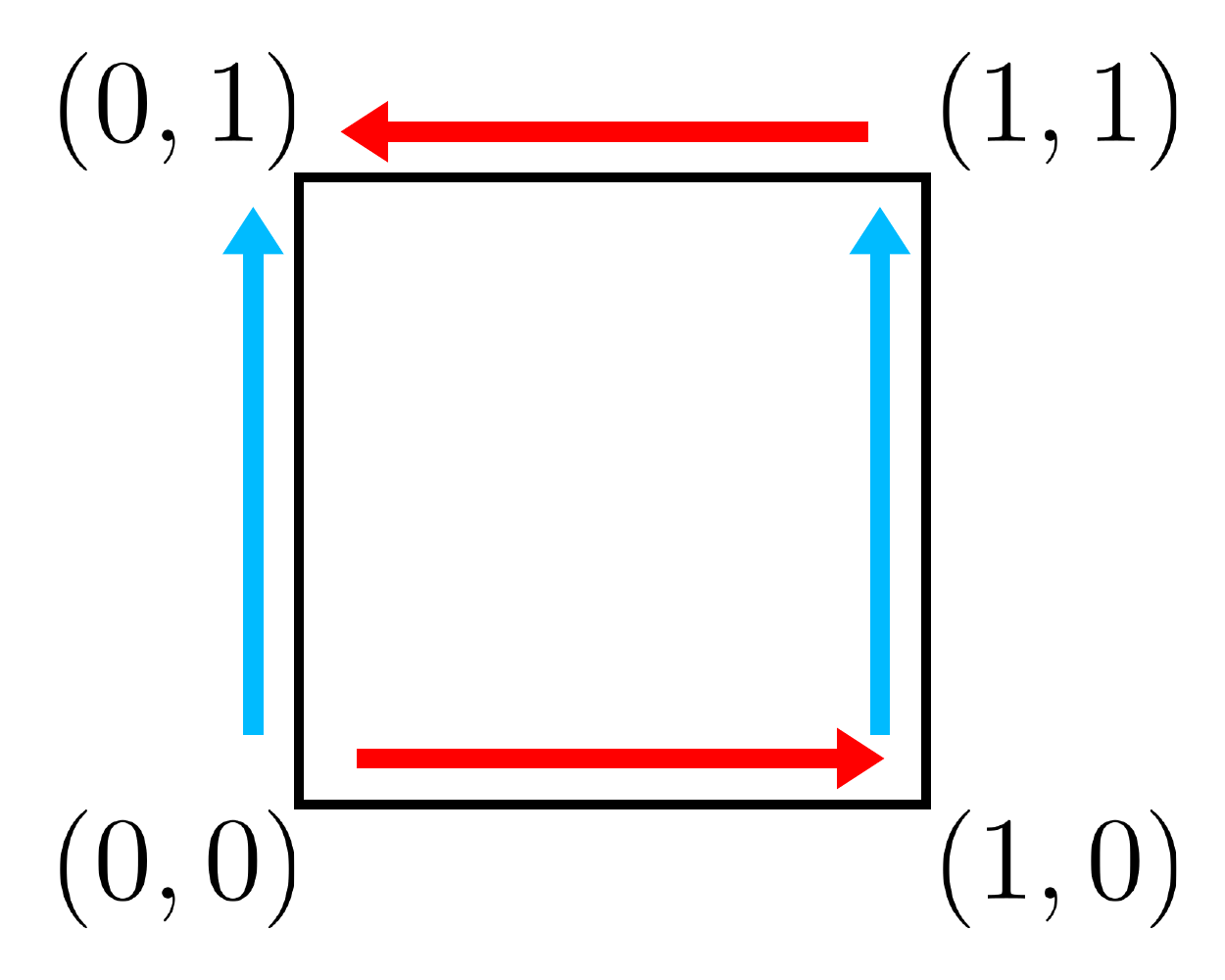} 
\end{minipage} &
\begin{tabular}[c]{@{}c@{}}$B = [b1,b2]$\\ $b1 = [1, 0]$\\ $b2 = [0, 2] $\end{tabular} & 
\begin{tabular}[c]{@{}c@{}}$B$-periodic\\ $(\ell_1,\mu_1)$-transflection\\ $\ell_1 \equiv (\frac{1}{2}, 0) + \lambda (0, 1)$
\\ $\mu_1 = 1$\end{tabular} &  
$V(z) = \mathcal{G}_{\ell_1}^{\mu_1}(\mathcal{P}_B(F))(z)$ \\ \hline

\begin{tabular}[c]{@{}c@{}}Projective\\ Plane\end{tabular} 
& 
\begin{minipage}{0.1\linewidth}
    \includegraphics[width=\linewidth]{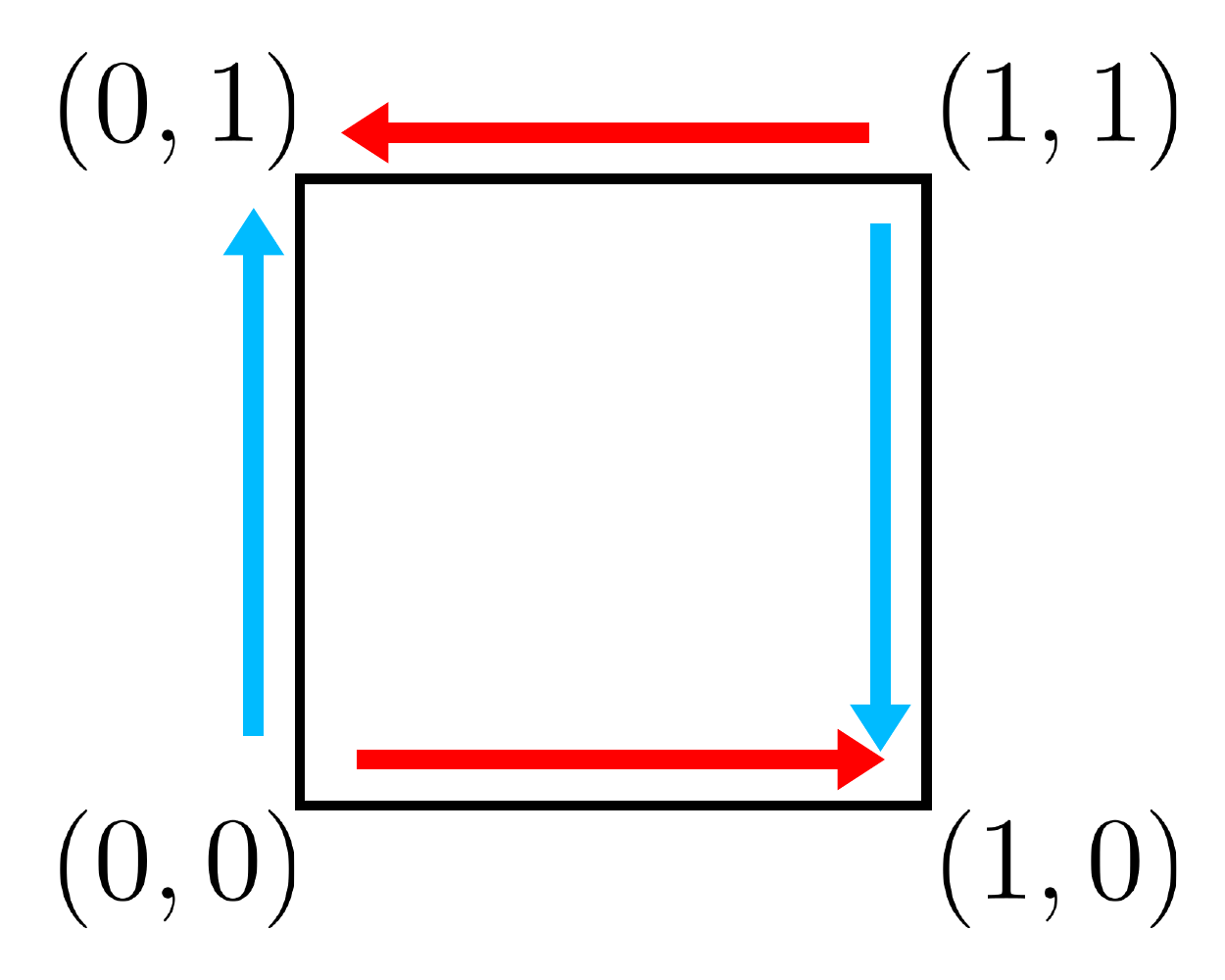} 
\end{minipage} &
\begin{tabular}[c]{@{}c@{}}$B = [b1,b2]$\\ $b1 = [2, 0]$\\ $b2 = [0, 2]$\end{tabular} & 
\begin{tabular}[c]{@{}c@{}}$B$-periodic\\ $(\ell_1,\mu_1)$-transflection\\ $(\ell_2,\mu_2)$-transflection\\ $ \ell_1 \equiv (0, \frac{1}{2}) + \lambda (1, 0)
$\\ $\ell_2 \equiv (\frac{1}{2}, 0) + \lambda (0, 1)
$\\ $\mu_1 = \mu_2 = 1$\end{tabular} 
 &  
$V(z) = \mathcal{G}_{\ell_1}^{\mu_1}(\mathcal{G}_{\ell_2}^{\mu_2}(\mathcal{P}_B(F)))(z)$ \\ \hline

\end{tabular}
}
\caption{Symmetrization for identification spaces}
\label{tab:IdentificationSymmetries}
\end{table*}

\subsection{Methodology}
Identification spaces arise from fundamental polygons subjected to specific edge identification constraints. Taking a square and applying different types of edge identification results in different types of spaces, such as torus, sphere, Klein bottle, and projective plane, as depicted in Fig. \ref{fig:Identification}. For each of these spaces, taking countable copies of the square and joining their edges according to the identification constraint results in a tiling pattern. This tiling pattern has internal symmetries such as periodicity, rotation, and transflection. Leveraging these symmetries, we introduce our Flow Symmetrization method to facilitate the construction of normalizing flows on these identification spaces.

% To do this, we exploit the inherent symmetries of the underlying manifold 
% employing a symmetrization process on the vector field of the flow, guided by the inherent symmetries of the underlying manifold. This process enforces the required geometric and topological constraints, ensuring the transformations are well-behaved across the manifold. It thus extends the applicability of normalizing flows to data residing on complex topological manifolds, providing a robust framework for density estimation in such settings. 
% We define \textit{Flow Symmetrization} as,

% \begin{quote}
%     Given an arbitrary vector field $\Vec{V}$, and a set of symmetry constraints $S$, apply the symmetrization process $\phi_S$, resulting in a vector field $\Vec{V}_S$ satisfying those required symmetries.
% \end{quote}

The core concept revolves around the behavior of points in \( \mathbb{R}^2 \) plane under the influence of a diffeomorphism. In this plane, a point subjected to a diffeomorphism transitions from an initial position to a final position. When edge identifications are applied to \( \mathbb{R}^2 \) plane, the space effectively collapses into a unit grid cell, where each cell is a replica of the other, possibly with rotations and reflections. Consequently, the motion of a point from one unit cell to another due to the diffeomorphism can be interpreted as a trajectory on the surface of the resulting non-Euclidean manifold formed through edge identification. Because the cells are replicas of each other, the point effectively always moves inside only one cell.

Let's consider the unit square cell \([0, 1]^2\), with one of the identified edges removed, and refer to it as the \textit{canonical square}. We denote the operation of projecting a point from an arbitrary cell to the \textit{canonical square} as the \textit{canonical projection} operation denoted by $\mathcal{P}: \mathbb{R}^2 \rightarrow [0,1]^2$. This projection is crucial as any point in a derived cell can be associated with the canonical square as every cell is a replica of each other due to the identification constraints. The canonical projection operation is pivotal in our approach to density estimation on identification spaces in the following way. We use normalizing flows on the Euclidean plane, which moves a point from the \textit{canonical square} to a point in some other cell. By projecting the point back to the canonical square, we ensure that the trajectories always remain inside the \textit{canonical square}. Moreover, the \textit{canonical square} being equivalent to the identification space ensures that the trajectory always remains inside the identification space. Thus, using a normalizing flow defined on the Euclidean space along with the \textit{canonical projection} operation, we design normalizing flows on the identification space. We now describe our approach in detail.

Let $z(t)$ be a finite continuous random variable having probability $p(z(t))$. Let the differential equations governing the change in $z$ with respect to time $t$ be given as
\begin{empheq}{align}
\frac{dz(t)}{dt}    & = V(z(t)) 
\end{empheq}
where $V$ is a vector field. This vector field induces a diffeomorphism under continuity assumptions on $V$ given as
\begin{empheq}{align}
D(z(t_0)) = z(t_1) = z(t_0) + \int_{t_0}^{t_1} V(z(t)) \,dt
\end{empheq}
Moreover, if $V$ is continuous, then the change in log probability can be obtained using the instantaneous change of variables \cite{chen2018neural} which is yet another differential equation given by
\begin{empheq}{align}
\frac{\partial \log p(z(t))}{\partial t} = -\operatorname{tr}\left(\frac{dV}{dz}\right)
\end{empheq}
Integrating this equation yields
\begin{empheq}{align}
\log p(z(t_1))=\log p(z(t_0)) -\int_{t_0}^{t_1} \operatorname{tr}\left(\frac{dV}{dz}\right)
\end{empheq}

Given the set \( X = \{x_1, x_2, \ldots, x_N\} \), we perform density estimation by minimizing the negative-log-likelihood function. Our loss function is given as

\begin{empheq}{align}
L = - \frac{1}{|X|} \sum_{i=1}^{N}\left( \log (q_0(\mathcal{P}(D_{\theta}(x_i)))) - \int_{t_0}^{t_1} \operatorname{tr}\left(\frac{dV}{dz}\right) \right)
\label{eq:NLL_lossfn}
\end{empheq}
 where $q_0$ is the template distribution (von Mises distribution in our case) and $D_{\theta}$ belongs to a parametric family of diffeomorphisms. The objective is to find the parameter $\theta$ for which the loss function in Eq. \ref{eq:NLL_lossfn} is minimized. We use the negative-log-likelihood as our objective function instead of the KL Divergence due to ease of optimization. 
 
 The vector field $V$ in Eq. \ref{eq:NLL_lossfn} is obtained by our Flow Symmetrization method. The symmetrization equation for each identification space is contained in Tab. \ref{tab:IdentificationSymmetries}. The identification constraints induce different types of symmetries. In the case of a torus, it is only periodicity, whereas in the case of a sphere, there is 4-fold rotational symmetry along with periodicity. In the case of the Klein bottle and projective plane, we also see the presence of transflection symmetry. Through the process of symmetrization, it is ensured that the vector flow field is identical in all the square cells. That is, the vector field inside the \textit{canonical square} is replicated elsewhere in all the square cells.

\begin{table}[!t]
\centering
\begin{tabular}{|c|c|c|c|}
\hline
 & 4-Gaussian & 6-Gaussian & 5x5 Checkerboard \\ \hline
Torus & $0.03$ & $0.45$ & $0.67$ \\ \hline
Sphere & $0.03$ & $0.47$ & $0.56$ \\ \hline
\end{tabular}
\caption{KL divergence between the target distribution and the distribution estimated by our method are displayed for the Torus and the Sphere on three different target distributions: 4-Gaussian, 6-Gaussian, and 5x5 Checkerboard.}
\label{tab:KLdivergence}
\end{table}

\subsection{Results}

\begin{figure}[!t]
  \centering
  \includegraphics[width=\linewidth]{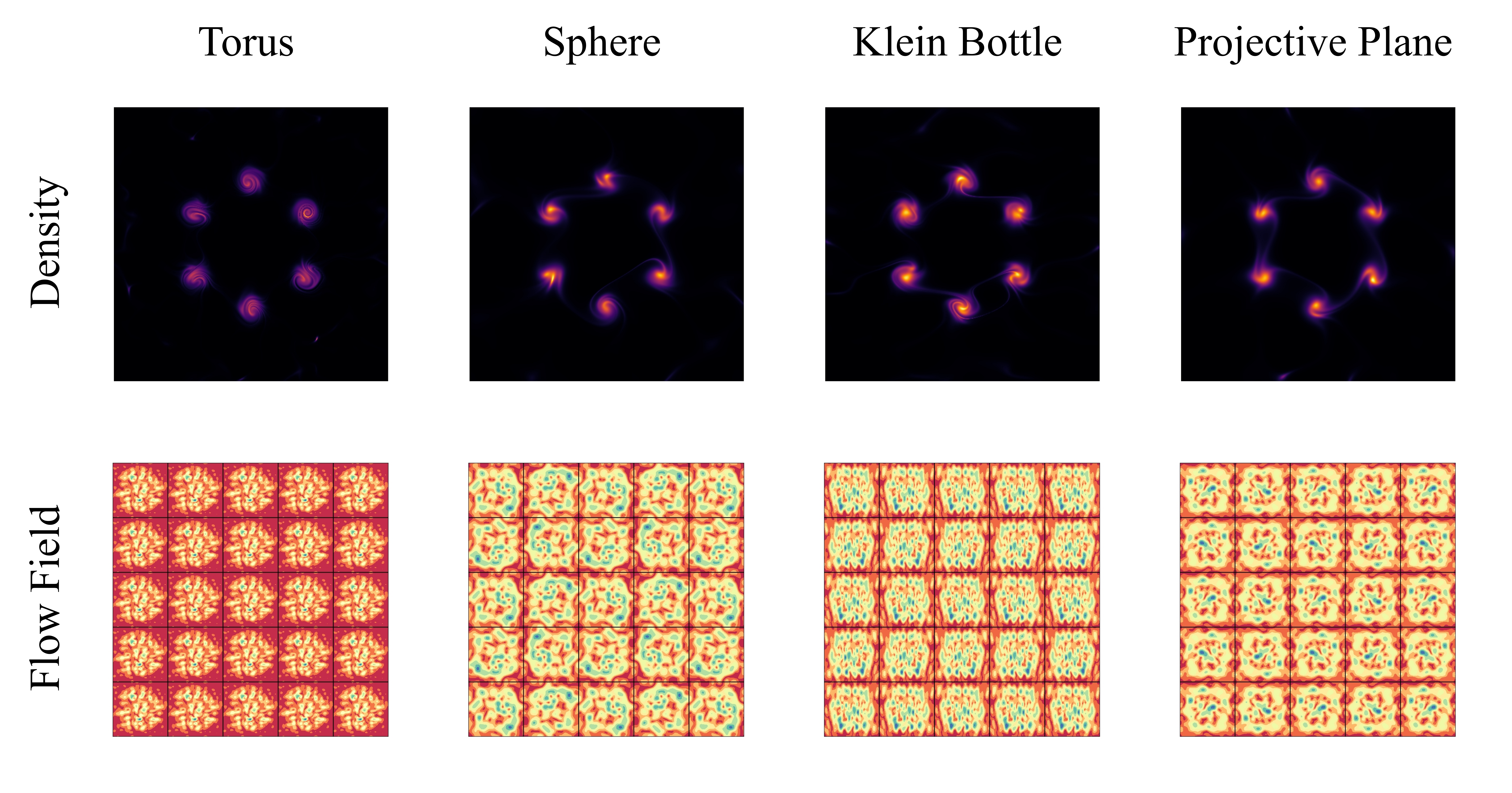}
  \caption{Result of Density Estimation for target distribution having a mixture of 6 Gaussians.}
  \label{fig:6Gaussian}
\end{figure}

\begin{figure}[!t]
  \centering
  \includegraphics[width=\linewidth]{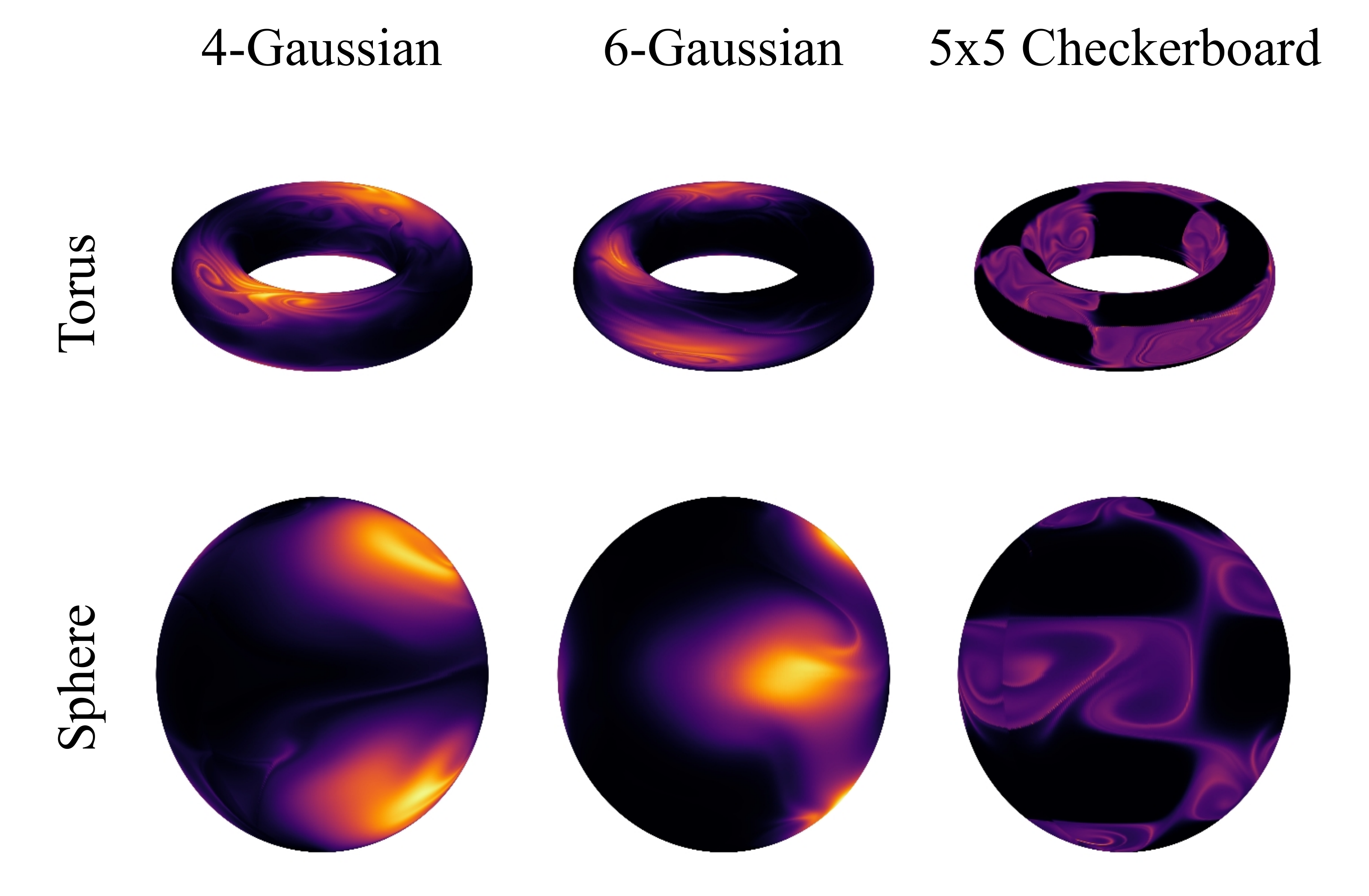}
  \caption{Visualization of the results of density estimation on torus and sphere for three different target distributions: 4-Gaussian, 6-Gaussian, and 5x5 Checkerboard.}
  \label{fig:SphereTorusDist}
\end{figure}

\begin{figure}[!t]
  \centering
  \includegraphics[width=\linewidth]{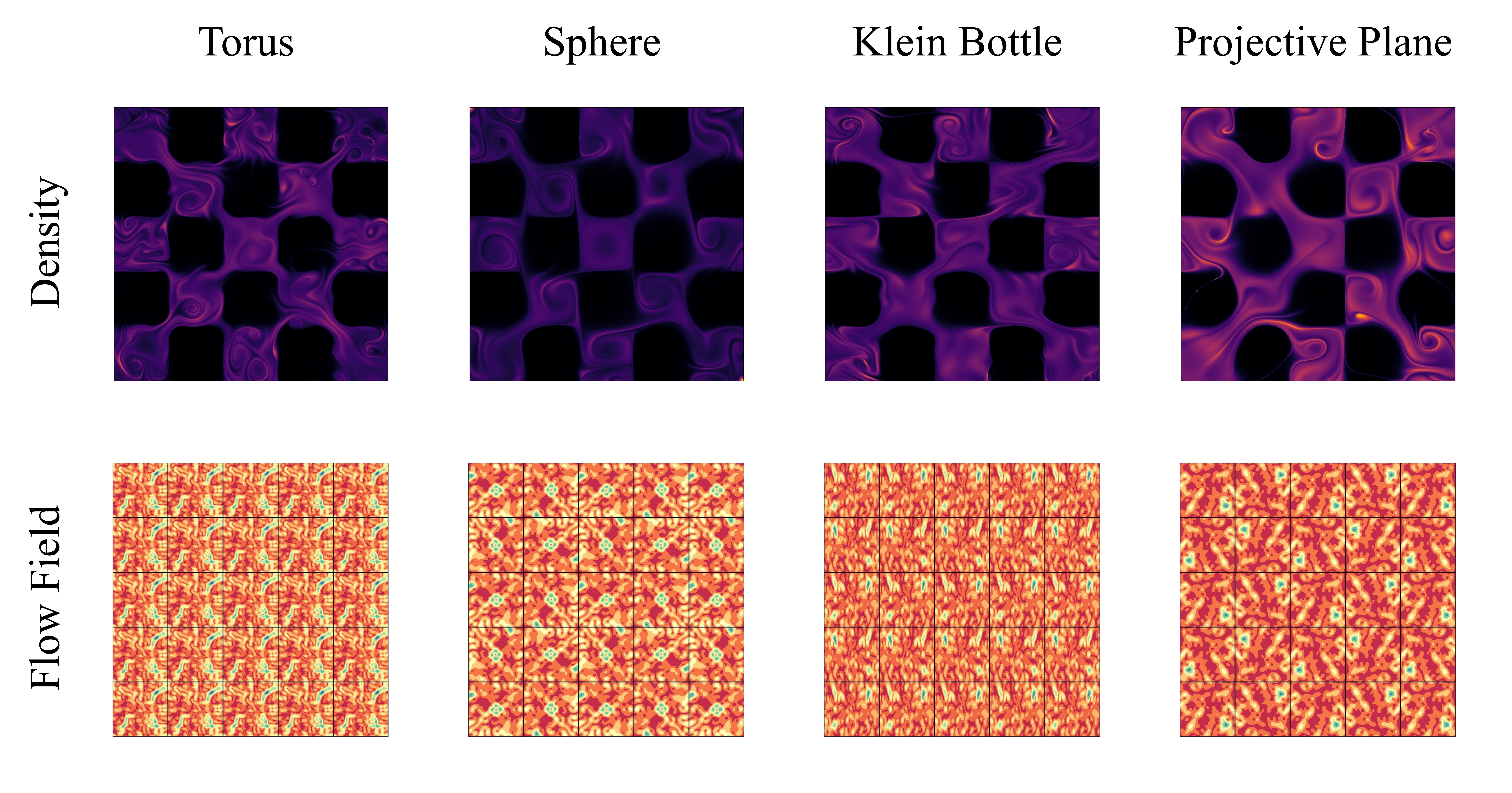}
  \caption{Results of Density Estimation for target distribution having 5x5 checkerboard pattern.}
  \label{fig:Checkerboard}
\end{figure}

In figures \ref{fig:6Gaussian} and \ref{fig:Checkerboard}, we present density estimation results for four different identification spaces: the torus, sphere, Klein bottle, and projective plane. Specifically, Figure \ref{fig:6Gaussian} illustrates the density estimation for a 6-Gaussian target distribution, while Figure \ref{fig:Checkerboard} focuses on a 5x5 checkerboard distribution. In these figures, black regions signify areas with low probability. The flow fields depicted in the figures represent the normalizing flows in \( \mathbb{R}^2 \), constructed in accordance with the symmetry constraints of the respective identification spaces. For example, the flow field associated with the Sphere clearly demonstrates \( C_4 \) rotational symmetry, as corroborated by Figure \ref{fig:Identification}. 
In Figure \ref{fig:SphereTorusDist} the result of density estimation for Sphere and Torus are visualized in 3D with the standard embedding for three different distributions: 4-Gaussian, 6-Gaussian, and 5x5 Checkerboard. The KL divergence between the target and the estimated distribution is depicted in Tab. \ref{tab:KLdivergence}.
These results demonstrate the effectiveness of our methodology in generating symmetric diffeomorphic flows on \( \mathbb{R}^2 \), thereby addressing the challenges of density estimation on identification spaces.

% target unknown distribution se points sample hoke aaye, and then we deform this template distribution, and log-likelihood ko maximize karne ke liye ek loss function aata hai USS mai. Template distribution ko diffeomorphism se pas karne ke baad woh distribution change ho jaata hai and -ve loglikelihood ko minimize karte hai. For making this on sphere we use lattice distribution.  

\section{Related Work}

Research has been conducted on the topics of Escherization, flow on manifold, and density estimation independently in the past. In this section, we provide a comprehensive overview of these works and highlight their limitations. We also present a unifying approach to these seemingly disparate domains, which enables us to solve them effectively and efficiently.

\label{sec:RelatedWork}

\textbf{Escherization:} \cite{kaplan2000escherization} proposed the generation of Escher tiles as an optimization problem and formulated the problem using a 6-point polygon to represent each isohedral tile. The optimization was performed using a simulated annealing algorithm that sifted through a parameterization space. This method was further enhanced by \cite{koizumi2011maximum}, who provided an analytical solution representing the goal and tile shapes as n-point polygons, thus linearizing the parameterization of tile shapes and reducing the problem to determining the maximum eigenvalue. Subsequent research by \cite{imahori2013local} extended this approach by proposing a local search algorithm for exploring alternative configurations. \cite{nagata2020efficient} introduced an efficient algorithm that comprehensively explores various templates in a reasonable computation time. However, despite this advancement, the \textit{n} points polygon method has a limitation: it does not incorporate shape deformation, limiting its ability to produce satisfactory results in complex goal-shape scenarios. To address this limitation, \cite{nagata2021escherization} proposed a new method based on the as-rigid-as-possible (ARAP) \cite{sorkine2007rigid} scheme. They create a mesh on the goal polygon, employ the ARAP-based distance function between goal shapes and input shape template, and incorporate this distance formulation in the EST (search algorithm).

% Letting \(\mathcal{D}\) denote the space of diffeomorphisms, and \(\mathcal Q\) be the shape space, one can use the transformation \(\pi: \mathcal{D} \to \mathcal Q\) to  \emph{project} a mathematical structure defined on diffeomorphisms to the shape space. With this method, one can, with just one modeling function, \(\mathcal{D}\), design many shape spaces, curved surfaces, images, density functions or measures, etc. 

\textbf{Diffeomorphisms in Shape Analysis}: Diffeomorphisms are invertible mappings with smooth forward and inverse transformations commonly used in shape analysis and nonrigid registration. The challenge of incorporating them into deep learning models has been a computational bottleneck. Hauberg et al. \cite{hauberg2016dreaming} attempted to address this by exploring data augmentation through learned class-dependent transformations using a generative model of diffeomorphisms. This approach demonstrated its efficacy in enhancing the performance of deep learning models. A significant advancement was made by Detlefsen et al. \cite{detlefsen2018deep}, who were among the first to successfully integrate diffeomorphisms into deep learning frameworks using continuous piecewise-affine velocity fields. Building on this, Gupta et al. \cite{gupta2020neural} introduced a shape auto-encoder for mesh deformations that leveraged Neural Ordinary Differential Equations (NODEs) to model shape spaces for genus-0 shapes. Sun et al. \cite{sun2022topologypreserving} further extended this by introducing Neural Diffeomorphic Flow (NDF), which also uses NODEs but focuses on preserving topological features during 3D reconstruction and registration. Most recently, Occupancy Flows by Mahjourian et al. \cite{mahjourian2022occupancy} and Neural Parts by Paschalidou et al. \cite{paschalidou2021neuralparts} have pushed the boundaries by learning continuous vector fields for 4D reconstruction and defining 3D primitives as homeomorphic mappings, respectively, allowing for more expressive and versatile shape representations.

% A vast body of work over the past decade has been dedicated to the definitions of shapes and shape spaces as mathematical objects and, therefore, their contributions and applications to domains in computer graphics, computer vision, etc. Letting $\mathcal{D}$ denote the space of diffeomorphisms, and $\mathcal Q$ be the shape space, one can use the transformation $\pi: \mathcal{D} \to \mathcal Q$ to  \emph{project} a mathematical structure defined on diffeomorphisms to the shape space. With this method, one can, with just one modeling function, $\mathcal{D}$, design many shape spaces, curved surfaces, images, density functions or measures, etc. Deep Learning-based methods for modeling diffeomorphisms were pioneered by \cite{detlefsen2018deep}, using continuous piecewise-affine velocity fields. Another work, on Occupancy Flows \cite{mahjourian2022occupancy}, seeks to learn a temporally and spatially continuous vector field that assigns a motion vector to every point in space and time in order to perform dense 4D reconstruction from images or sparse point clouds. However, most work in this area focuses on the representation and parameterization of surfaces, and little has gone into exploiting constrained diffeomorphisms for the analysis of shapes.
\textbf{Normalizing flows as diffeomorphisms}: With the introduction of Neural Ordinary Differential Equations (NODE) solvers \cite{chen2018neural}, numerical integration over the flow, and hence, gradient-based optimization over the space of diffeomorphisms, became feasible using normalizing flows. Normalizing flows have recently emerged as front-runners in the field of shape representation and shape analysis, primarily because they have been proven to model diffeomorphisms effectively \cite{lei2022cadex} \cite{sun2022topology}\cite{gupta2020neural} \cite{ma2022cortexode} \cite{jiang2020shapeflow} \cite{dummer2023rsa}. Neural flow techniques have also been employed to generate diffeomorphisms for modeling shape spaces for genus-0 shapes through a shape auto-encoder for mesh deformations \cite{gupta2020neural}. Diffeomorphisms, primarily represented through \emph{flows}, have been utilized for shape reconstruction while preserving topological properties \cite{sun2022topologypreserving}. The work by Niemeyer et al. \cite{niemeyer2019occupancy} introduces Occupancy Flow, a spatio-temporal representation that enables dense 4D reconstruction from images or sparse point clouds, providing a versatile approach for various spatio-temporal reconstruction tasks. The model by Grathwohl et al. \cite{grathwohl2019ffjord} extends the capabilities of normalizing flows by introducing free-form Jacobian of Reversible Dynamics (FFJORD), which allows for more flexible transformations and has applications in generative modeling.

\textbf{Density Estimation and Flows on Manifolds}: The work began by constructing flows on the Riemannian manifolds that are diffeomorphic to Euclidean space \cite{gemici2016normalizing}. The paper \cite{rezende2020normalizing} introduced normalizing flows for tori and spheres, and \cite{bose2020latent} extended this approach to hyperbolic spaces. Concurrently, \cite{lou2020neural}, \cite{mathieu2020riemannian}, and \cite{falorsi2020neural} developed flows on Riemannian Manifolds by extending Neural ODEs. \cite{katsman2021equivariant} developed symmetry-invariant distributions on arbitrary manifolds via equivariant manifold
flows. By explicitly incorporating symmetry within manifold flows and employing density estimation techniques, they perform better than general manifold flows in scenarios with inherent symmetry.

\section{Conclusion and Future Work}

We introduce Flow Symmetrization, a novel method to represent a parametric family of diffeomorphisms that satisfy identification constraints. The symmetries that arise due to identification constraints are periodicity, rotation equivariance, and transflection equivariance. In our method, the underlying vector field of the diffeomorphism is enforced to be symmetric through a process of symmetrization. We apply this method to the problem of Escherization on the plane and for density estimation on the identification spaces.

Exploring other loss functions for comparing the tile shape with the target shape in the Escherization problem would be an interesting future direction. In density estimation, we currently show results on 2-dimensional identification spaces. However, the idea of our method applies in general to any dimension. Thus, a future goal is to extend our density estimation framework to work on identification spaces of higher dimensions, where instead of edge identification in polygons we would have facet identification in polyhedra. Another aspect is the memory requirement in our approach, which scales with the maximum allowed frequency in the Fourier representation. Making our method memory efficient is one of our future goals as this would allow the accommodation of higher frequency values, thus facilitating the search over highly complex tile shapes. This would also enable us to solve other tiling-related shape search problems.

Utilizing diffeomorphisms with identification constraints, we identify and exploit a subtle link between two traditionally distinct problems: Escherization and density estimation. Our method's adaptability ensures not only a seamless integration across domains but also achieves impressive results, underscoring the importance of our Flow Symmetrization method.

\bibliographystyle{eg-alpha-doi} 
\bibliography{egbibsample}

\end{document}